\documentclass[prd,aps,showpacs,nofootinbib,preprintnumbers,onecolumn]{revtex4}
\usepackage{amsfonts,amsmath,amssymb,bm,dsfont,eucal}
\usepackage{multirow}
\usepackage[english]{babel}
\usepackage[latin1]{inputenc}
\usepackage[T1]{fontenc}
\usepackage{ae,aecompl}
\usepackage{graphicx,color}
\usepackage{graphicx,graphics,epsfig,epstopdf,subfigure,color,psfrag}
\usepackage{hyperref}

\newcommand{\spur}[1]{\not\! #1 \,}
\newcommand{\be}{\begin{equation}}
\newcommand{\ee}{\end{equation}}
\newcommand{\bea}{\begin{eqnarray}}
\newcommand{\eea}{\end{eqnarray}}
\newcommand{\nn}{\nonumber}
\newcommand{\dd}{\displaystyle}

\begin{document}

\preprint{BARI-TH/651-12}
\title{New  meson spectroscopy with  open charm and beauty}
\author{P. Colangelo$^a$,  F.  De Fazio$^a$, F. Giannuzzi$^{a,b}$ and S. Nicotri$^{a,b}$}
\affiliation{$^a$Istituto Nazionale di Fisica Nucleare, Sezione di Bari, Italy\\$^b$Dipartimento di Fisica,
 Universit$\grave{a}$ degli Studi di Bari, Italy}

\begin{abstract}
All the available experimental information on open charm and beauty mesons is used to classify   the observed states in heavy quark doublets. The masses of some of the  still unobserved states are predicted, in particular in the beauty sector.
 Adopting an effective Lagrangian approach based  on the heavy quark  and  chiral symmetry,  individual decay rates and ratios of branching fractions are computed, with results useful to assign the  quantum numbers to  recently observed charmed states which still need to be properly classified. Implications and predictions for the corresponding beauty mesons are provided.  The experimental results  are already copious, and are expected to grow up thanks to the experiments at the LHC and to  the future high-luminosity flavour and $p-\bar p$ facilities.

\end{abstract}

\maketitle

\section{Introduction}
The analysis of hadrons containing a single heavy quark can be afforded in a  predictive framework exploiting the heavy quark (HQ) limit, which is  formalized in the heavy quark effective theory (HQET) \cite{rev0}.  This is an effective theory formulated for $N_f$ heavy quarks $Q$ with mass $m_Q \gg \Lambda_{QCD}$ (a condition generally denoted as  the  $m_Q \to \infty$ limit),  keeping the four-velocity of $Q$ fixed. The theory displays heavy quark spin-flavour symmetries, i.e. invariance under $SU(2N_f)$ transformations. Within this framework it is possible to describe  several heavy hadron  properties, the prime example being the relations among semileptonic transition form factors in weak heavy hadron matrix elements \cite{Isgur:1989ed}. In this paper we consider another sector in which the heavy quark limit provides powerful predictions,  the  heavy meson spectroscopy \cite{Isgur:1991wq}.  The renewed interest in this subject is due to the remarkable number of  resonances discovered in the recent years,  a set of states that will presumably become richer with the data taking of the  experiments at the CERN LHC, as well as of the new high-luminosity flavour factories and $p - \bar p$ facilities.

The classification of heavy $Q{\bar q}$ mesons ($q$ being a light quark) in the HQ limit relies on the decoupling of the heavy quark spin from the spin of  light degrees of freedom,  the light antiquark and the gluons. The spin $s_Q$ of the heavy quark and the total angular momentum of the light degrees of freedom $s_\ell$ are separately conserved in strong interaction processes. Heavy mesons can therefore be classified according to the value of $s_\ell$, and can be collected in doublets;  the two states of each doublet have total spin  $J=s_\ell \pm {1 \over 2}$ and parity $P=(-1)^{\ell +1}$, with $\ell$  the orbital angular momentum of the light degrees of freedom and ${\vec s}_\ell={\vec \ell}+ {\vec s}_q$ ($s_q$ is the light antiquark spin).  We  refer to the two states in the same doublet as the spin partners.
Since the properties of hadron states are independent of the spin and  flavour of the heavy quark, the two states within each doublet are degenerate in the HQ limit and, due to flavour symmetry, the properties of the states in a  doublet can be related to those of the corresponding states that differ for the flavour of the heavy quark.
In practice, HQET turns out to be a theory of charmed and beauty hadrons,  the top quark is too heavy to hadronize before decaying.

In the following we focus on  the meson doublets corresponding to $\ell=0,1,2$ (in the quark model they are  referred to  as $s-$, $p-$ and $d-$wave states), we discuss their properties in the  HQ limit and include corrections  to the degeneracy condition within each doublet. This allows us to study how the observed charmed and beauty mesons fit in the theoretical classification. Furthermore, using the available data in the charm sector,  the properties of the corresponding beauty mesons, if not yet observed, can be predicted.

Since several recently discovered states  require a proper  classification, it is  useful to analyze strong decays of heavy mesons to light pseudoscalar mesons, because the decay rates of these processes depend on the quantum numbers of the decaying resonances. For such a purpose, we exploit an effective Lagrangian approach in which the heavy quark doublets are represented by effective fields, while the octet of light  pseudo Goldstone mesons is grouped  in a single field.  The Lagrangian terms describing the strong decays of a heavy meson with the emission of a light pseudoscalar meson are invariant both under heavy quark spin-flavour transformations and chiral transformations  of the light pseudo Goldstone boson fields.

In the exact HQ limit for the heavy meson doublets, considering the strong decays with  a light pseudoscalar meson in the final state, as a consequence of the spin flavour symmetries,
 the following properties are expected \cite{Isgur:1991wq}:
\begin{itemize}
\item The two states within a  doublet are degenerate in mass;
\item The two states within a  doublet have the same full width;
\item The sum of the partial widths of a state in a doublet to another heavy state in another doublet with emission of a light meson is the same for the two states of a doublet;
\item Spin symmetry predicts the ratios of partial decay widths for a given state;
\item Partial decay widths are independent of the heavy quark flavour;
\item Mass splittings among the different doublets are also independent of the heavy quark flavour.
\end{itemize}
 In the following Sections  we analyze the complete set of the established and newly observed open charm and beauty mesons, testing   the above properties and  discussing the possible deviations,
with the aim of a unique description for all the experimental findings. On the basis of such a description, several predictions for still unobserved resonances, in particular in the beauty sector, will be obtained.

\section{Heavy meson doublets: an overview}
We  discuss the states of the doublets corresponding to $\ell=0,1,2$.
The lowest lying $Q \bar q$ mesons correspond to $\ell=0$, then
$ s_\ell^P={1 \over 2}^-$;
 this  doublet consists of  two states with spin-parity
$J^P=(0^-,1^-)$ that we denote as $(P,P^*)$. For $\ell=1$ it could
be either $ s_\ell^P={1 \over 2}^+$ or $ s_\ell^P={3 \over 2}^+$.
The  two corresponding doublets have $J^P=(0^+,1^+)$ and
$J^P=(1^+,2^+)$.  We denote the members of the
$J^P_{s_\ell}=(0^+,1^+)_{{1/2}}$ doublet as
$(P^*_{0},P_{1}^\prime)$,
 and those of the
$J^P_{s_\ell}=(1^+,2^+)_{3/2}$  doublet as $(P_{1},P^*_{2})$.
$\ell=2$ corresponds to either $s_\ell^P={3 \over 2}^-$ or $s_\ell^P={5 \over 2}^-$; the states belonging to such doublets are denoted as $(P_1^*,P_2)$ and $(P_2^{\prime *}, P_3)$, respectively. We use an analogous notation for the radial excitations of these states with radial quantum number $n=2$,   distinguishing their fields by a tilde ($\tilde P$, $\tilde P^*$, ...).

The expressions for  the effective fields describing the various doublets in the HQ limit are collected below: $H_a$ ($a=u,d,s$ a light flavour index) corresponds to $s_\ell^P={ 1 \over 2}^-$; $S_a$ and $T_a$  to $ s_\ell^P={1 \over 2}^+$ and $s_\ell^P={3 \over 2}^+$, respectively; $X_a$ describes the doublet with $ s_\ell^P={3 \over 2}^-$,  and $X_a^\prime$  the $s_\ell^P={5 \over 2}^-$ doublet:
\bea
H_a & =& \frac{1+{\rlap{v}/}}{2}[P_{a\mu}^*\gamma^\mu-P_a\gamma_5]  \label{neg} \nn  \\
S_a &=& \frac{1+{\rlap{v}/}}{2} \left[P_{1a}^{\prime \mu}\gamma_\mu\gamma_5-P_{0a}^*\right]   \nn \\
T_a^\mu &=&\frac{1+{\rlap{v}/}}{2} \Bigg\{ P^{\mu\nu}_{2a}
\gamma_\nu - P_{1a\nu} \sqrt{3 \over 2} \gamma_5 \left[ g^{\mu
\nu}-{1 \over 3} \gamma^\nu (\gamma^\mu-v^\mu) \right]
\Bigg\}  \hspace*{1.2cm} \label{pos2} \\
X_a^\mu &=&\frac{1+{\rlap{v}/}}{2} \Bigg\{ P^{*\mu\nu}_{2a}
\gamma_5 \gamma_\nu -P^{\prime *}_{1a\nu} \sqrt{3 \over 2}  \left[
g^{\mu \nu}-{1 \over 3} \gamma^\nu (\gamma^\mu+v^\mu) \right]
\Bigg\}   \nn   \\
X_a^{\prime \mu\nu} &=&\frac{1+{\rlap{v}/}}{2} \Bigg\{
P^{\mu\nu\sigma}_{3a} \gamma_\sigma -P^{*'\alpha\beta}_{2a}
\sqrt{5 \over 3} \gamma_5 \Bigg[ g^\mu_\alpha g^\nu_\beta - {1
\over 5} \gamma_\alpha g^\nu_\beta (\gamma^\mu-v^\mu) -  {1 \over
5} \gamma_\beta g^\mu_\alpha (\gamma^\nu-v^\nu) \Bigg] \Bigg\} \nn
\,\,.\eea
The various operators in Eq.(\ref{neg})
annihilate mesons of four velocity $v$, which is conserved in strong
interaction processes; they  include a factor $\sqrt{m_Q}$ and have dimension $3/2$.
The octet of light pseudoscalar mesons is
introduced through the definitions $\displaystyle \xi=e^{i {\CMcal M}
\over f_\pi}$ and $\Sigma=\xi^2$,  with  the matrix ${\CMcal M}$
incorporating the  $\pi, K$ and $\eta$ fields ($f_{\pi}=132 \; $ MeV):
\begin{equation}
{\CMcal M}= \left(\begin{array}{ccc}
\sqrt{\frac{1}{2}}\pi^0+\sqrt{\frac{1}{6}}\eta & \pi^+ & K^+\\
\pi^- & -\sqrt{\frac{1}{2}}\pi^0+\sqrt{\frac{1}{6}}\eta & K^0\\
K^- & {\bar K}^0 &-\sqrt{\frac{2}{3}}\eta
\end{array}\right) \,\,\,\, . \label{pseudo-octet}
\end{equation}
With these fields an effective Lagrangian can be written, invariant under heavy quark spin-flavour  and light quark chiral transformations \cite{hqet_chir,wise_book}.
The   kinetic  terms of the heavy meson doublets  and  of the $\Sigma$  field read:
 \begin{eqnarray} {\CMcal L} &=& i\; Tr[ {\bar H}_b v^\mu
D_{\mu ba}  H_a ]  + \frac{f_\pi^2}{8}
Tr[\partial^\mu\Sigma\partial_\mu \Sigma^\dagger] \nn \\ &+&
Tr[ {\bar S}_b \;( i \; v^\mu D_{\mu ba} \; - \; \delta_{ba} \;
\Delta_S)  S_a ]
+   Tr[ {\bar T}_b^\alpha \;( i \; v^\mu D_{\mu ba} \; - \;
\delta_{ba} \; \Delta_T)  T_{a \alpha} ]   \label{L}\\
&+&
Tr[ {\bar X}_b^\alpha \;( i \; v^\mu D_{\mu ba} \; - \; \delta_{ba} \;
\Delta_X)
 X_{a \alpha} ]
+   Tr[ {\bar X}_b^{ \prime \alpha \beta} \;( i \; v^\mu D_{\mu ba} \; - \;
\delta_{ba} \; \Delta_{X^\prime})  X_{a \alpha \beta}^\prime ] \,\,\, . \nonumber
\end{eqnarray}
Such terms  involve the operators $D$ and $\CMcal A$:
\begin{eqnarray}
D_{\mu ba}&=&-\delta_{ba}\partial_\mu+{\CMcal V}_{\mu ba}
=-\delta_{ba}\partial_\mu+\frac{1}{2}\left(\xi^\dagger\partial_\mu
\xi
+\xi\partial_\mu \xi^\dagger\right)_{ba}\\
{\CMcal A}_{\mu ba}&=&\frac{i}{2}\left(\xi^\dagger\partial_\mu
\xi-\xi
\partial_\mu \xi^\dagger\right)_{ba} \; .
\end{eqnarray}
The mass parameters $\Delta_F$ (with $F=S,\,T,\,X,\,X^\prime$)
represent the mass splittings between  the  higher mass doublets and the lowest lying  doublet described by the field $H$;
 they  can be expressed in terms of the spin-averaged masses of the doublets:
 \be
 \Delta_F= \overline M_F - \overline M_H
 \ee
 with
\bea {\overline M}_H& =& {3 M_{P^*}+M_P  \over 4} \nn \\
{\overline M}_S &=& {3 M_{P^\prime_1}+M_{P_0^*} \over 4} \nn \\
{\overline M}_T &=& {5 M_{P^*_2}+3M_{P_1} \over 8}  \label{masse-medie} \\
{\overline M}_X &=& {5 M_{P_2}+3M_{P_1^*} \over 8} \nn \\
{\overline M}_{X^\prime} &=& {7 M_{P_3}+5M_{P_2^{\prime *}} \over 12} \nn\,\, .
\eea

Corrections to the heavy quark limit  are represented by  symmetry breaking
terms  suppressed by increasing powers of  the inverse heavy quark mass $m_Q$
\cite{Falk:1995th}. In particular, the mass degeneracy between the members of the
meson doublets is broken by the  terms:
\bea
{\CMcal L}_{1/m_{Q}}&=&{1 \over 2 m_{Q}} \Big\{  \lambda_H Tr [{\bar
H}_{a} \sigma^{\mu \nu} H_{a} \sigma_{\mu \nu}]+\lambda_S Tr
[{\bar S}_{a} \sigma^{\mu \nu} S_{a} \sigma_{\mu \nu}]+\lambda_T
Tr [{\bar T}^\alpha_{a}
\sigma^{\mu \nu} T^\alpha_{a} \sigma_{\mu \nu}] \nn \\
&&+\lambda_X Tr [{\bar X}_{a}^\alpha \sigma^{\mu \nu} X_{a \alpha}
\sigma_{\mu \nu}] +\lambda_{X^\prime} Tr [{\bar X}^{\prime \alpha
\beta}_{a} \sigma^{\mu \nu} X^{\prime \alpha \beta}_{a}
\sigma_{\mu \nu}] \Big\} \,\,\,\, , \label{mass-viol} \eea
with the constants $\lambda_H$, $\lambda_S$, $\lambda_T$, $\lambda_X$ and
$\lambda_{X^\prime}$  connected  to the hyperfine  splitting in each doublet:
\bea
\lambda_H &=& {1 \over 8} \left( M_{P^*}^2-M_P^2 \right)
\nonumber \\
\lambda_S &=& {1 \over 8} \left(M_{P^\prime_1}^2-M_{P_0^*}^2 \right)
\nonumber \\
\lambda_T &=& {3 \over16} \left( M_{P^*_2}^2-M_{P_1}^2 \right) \label{lambdat}\\
\lambda_X&=&{3 \over16} \left( M_{P_2}^2-M_{P_1^*}^2 \right)
\nonumber  \\
\lambda_{X^\prime}&=&{5 \over 24} \left( M_{P_3}^2-M_{P_2^{\prime*}}^2 \right) \,\,\, . \nonumber
\eea

\begin{table*}[b]
\centering \caption{Observed open charm and open beauty  mesons, classified in HQ doublets. States with uncertain  assignment are indicated with $\star$  and  classified according to the  scheme proposed in this study.}\label{charm}
\begin{tabular}{|c   |   c || c c |c c|| c c | c c | }\hline
doublet  \,\,$s_\ell^P$& \,\,\,$J^P$ \,\,\,& \,\,$c{\bar q}$ \, (n=1) \,\,& \,\,$c{\bar q}$ \,(n=2)\,\, & \,\, $c{\bar s}$ \, (n=1) \,\,& \,\, $c{\bar s}$ \, (n=2) \,\,
& \,\,$b{\bar q}$ \, (n=1) \,\,& \,\,$b{\bar q}$ \,(n=2) & \,\, $b{\bar s}$ \, (n=1) \,& \,\, $b{\bar s}$ \, (n=2) \\
\hline \hline
\multirow{2}{*}{\,\,\,\,\,\,\,\,$H$ \,\,\,\,\,\,\,\,\, $\frac{1}{2}^- $  }&  $0^-$ & $ D(1869)$ &
$D(2550)$ $\star$  & $D_s(1968)$  & & $ B(5279)$ & & $ B_s(5366)$ &
\\  \cline{2-10}
   & $1^-$ & $ D^*(2010)$ & $D^* (2600)$ $\star$  & $D_s^*(2112)$  & $D_{s1}^{*}(2700)$ & $ B^*(5325)$ & & $ B_s^*(5415)$ &
\\ \hline \hline
\multirow{2}{*}{\,\,\,\,\,\,\,\,$S$ \,\,\,\,\,\,\,\,\, $\frac{1}{2}^+ $ }&   $0^+$ & $ D^*_0(2400)$ &     & $D_{s0}^*(2317)$  & & & & &
\\ \cline{2-10}
  & $1^+$ &   $D^{ \prime}_1(2430)$  & & $D_{s1}^\prime(2460)$  & $D_{sJ}(3040)$ $\star$ & & & &
\\ \hline \hline
\multirow{2}{*}{\,\,\,\,\,\,\,\,$T$ \,\,\,\,\,\,\,\,\, $\frac{3}{2}^+ $ }   & $1^+$ & $D_1(2420)$  & & $D_{s1}(2536)$  & $D_{sJ}(3040)$ $\star$ & $ B_1(5721)$ & & $ B_{s1}(5830)$ &
\\ \cline{2-10}
   & $2^+$ &   $D^*_2(2460)$ & & $D_{s2}^*(2573)$  & & $ B_2^*(5747)$ & & $ B_{s2}^*(5840)$ &
\\ \hline \hline
\multirow{2}{*}{\,\,\,\,\,\,\,\,$X$  \,\,\,\,\,\,\,\,\, $\frac{3}{2}^- $ }   & $1^-$ &     & &   & & & & &
\\ \cline{2-10}
   & $2^-$ &     & & & & & & &
\\ \hline \hline
  \multirow{2}{*}{\,\,\,\,\,\,\,\,$X^\prime$ \,\,\,\,\,\,\,\,\, $\frac{5}{2}^- $ }&   $2^-$ &   $D(2750)$ $\star$  & &  & & & & &
\\ \cline{2-10}
   & $3^-$ &  $D(2760)$ $\star$ & &  $D_{sJ}(2860)$ $\star$ & & & & &
\\ \hline
  \end{tabular}
\end{table*}

In Table \ref{charm}  we collect the  observed charmed $c{\bar q}$ and $c{\bar s}$, and beauty $b{\bar q}$ and  $b{\bar s}$ (with $q=u,d$) mesons,   with a classification  established within the heavy quark doublet  scheme. In the Table
we  also include  states not yet classified: Their position reflects our proposed  assignment of their quantum numbers, and to indicate this (still unsettled) situation we put a $\star$ mark in correspondence of such mesons.  Alternative classifications  will also be discussed in the following.

The states belonging to the lowest   $s_\ell^P={1 \over 2}^-$  and $n=1$ doublet are well recognized,  hence our discussion mainly concerns the excited doublets with either $\ell ={1 \over 2}^+, {3 \over 2}^{\pm}, \dots$, or $n > 1$. In Tables \ref{charm-pdg} and \ref{beauty-pdg} we collect the values of the masses and widths of such resonances  as reported by the Particle Data Group  (PDG) \cite{pdg}. For the masses of the states of the lightest doublet, as well as for the masses of the light pseudoscalar mesons,  we refer to  the PDG  values.

\begin{table*}[b]
\centering \caption{Measured mass and width of the observed excited  mesons with open charm. All the results are from the PDG \cite{pdg}, excluding the data concerning  $D^{*0,+}(2600)$, $D^{0}(2750)$ and $D^{*0,+}(2760)$
which are BaBar measurements  \cite{delAmoSanchez:2010vq}; the widths of  $D^{*+}(2600)$ and  $D^{*+}(2600)$ are kept fixed in the experimental analysis  \cite{delAmoSanchez:2010vq}.  The quoted bounds are at $95\%$ CL.}\label{charm-pdg}
\begin{tabular}{|c  c c | c c c | }\hline
$c{\bar q}$  &   mass (MeV)  & $\Gamma$ (MeV) & $c{\bar s}$  & mass (MeV)\hskip 1 cm  &  $\Gamma$ (MeV)  \\ \hline
$D_0^{*0}(2400)$ & $2318 \pm 29$ & $267 \pm 40$ & & & \\\hline
$D_0^{*\pm}(2400)$ & $ 2403 \pm 14 \pm 35$ & $283 \pm 24 \pm 34$ & $D_{s0}^{*}(2317)$ & $2317.8 \pm 0.6 $ & $<3.8$ \\ \hline
$D_{1}^{\prime 0 }(2430)$ & $ 2427 \pm 26 \pm 25$ & $384 \pm^{107}_{75}  \pm 74$ &  & & \\\hline
& & & $D_{s1}^{\prime}(2460)$ & $2459.6 \pm 0.6$ & $<3.5$ \\\hline \hline
$D_1^0(2420)$ & $2421.3 \pm 0.6$ & $27.1 \pm 2.7$ & && \\\hline
$D_1^\pm(2420)$ & $ 2423.4 \pm 3.1$ &  $26 \pm 6 $ & $D_{s1}(2536)$ & $ 2535.12 \pm 0.13$ & $0.92\pm0.03\pm0.04$\\\hline
$D_2^{*0}(2460)$ & $2462.6 \pm 0.7$ & $ 49.0 \pm 1.4$ &&& \\\hline
$D_2^{*\pm}(2460) $ & $2464.4 \pm 1.9$ & $37 \pm 6$ &
$D_{s2}^{*}(2573)$ & $2572.6 \pm 0.9$ & $20 \pm 5$ \\\hline \hline
$D^0(2550) $ & $\,\,\,\, 2539.4 \pm 4.5 \pm 6.8\,\,\,\,\, $ & $130 \pm 12 \pm 13 $ & && \\\hline
$D^{*0}(2600)$ & $2608.7 \pm 2.4 \pm 2.5$ & $ 93 \pm 6 \pm 13 $ &&& \\\hline
$D^{*+}(2600)$ & $2621.3 \pm 3.7 \pm 4.2$ & $93$ (fixed) & $D_{s1}^*(2700)$ & $2709 \pm^9_6$ & $125 \pm 30$ \\\hline \hline
$D^0(2750)$ & $ 2752.4 \pm 1.7 \pm 2.7$ & $71 \pm 6 \pm 11$ &&& \\\hline
$D^{*0}(2760)$ & $ 2763.3 \pm 2.3 \pm 2.3$ & $60.9  \pm 5.1 \pm 3.6$ &&& \\\hline
$\,\,\,\, D^{*+}(2760)\,\,\,\, $ & $ 2769.7 \pm 3.8 \pm 1.5$ & $60.9$ (fixed) & $\,\,\,\, D_{sJ}(2860)\,\,\,\, $ & $2862 \pm 2 \pm^5_2$  & $48 \pm 3 \pm 6$ \\\hline \hline
&&& $D_{sJ}(3040)$ & $3044 \pm 8
\pm^{30}_{5}$ & $\,\,\,\, 239 \pm 35 \pm^{46}_{42}\,\,\,\, $
 \\ \hline
\end{tabular}
\end{table*}

Let us examine the various entries in  Table \ref{charm}.
The  $s_\ell^P={3 \over 2}^+$ charmed doublets are filled by the states $(D_1(2420), \,D_2^*(2460))$ and $(D_{s1}(2536),\,D_{s2}^*(2573))$ in the non-strange and strange case, respectively.
Considering their  widths  in Table \ref{charm-pdg}, it can be noticed that
these states are quite narrow,  in accordance with the expectation, since their strong decays occur in $d$-wave:  the widths of  strong decays to a light pseudoscalar meson of momentum $\vec p$ are proportional to ${|\vec p |}^{2 \ell +1}$,  with $\ell$ the angular momentum transferred in the decay. In these modes $|\vec p|$ is small,  and highest values of $\ell$ correspond to  the largest  suppression of the  decay rate.

The states $(D_0^*(2400), \, D_1^\prime(2430))$ and $(D_{s0}^*(2317), \, D_{s1}^\prime(2460))$ can be  identified with the members of the $s_\ell^P={1 \over 2}^+$ charm doublet, although they present  puzzling features.
The  non strange states follow the expectation of being broad,   their strong decays occurring in $s$-wave. Evidences of $c{\bar q}$ broad states were   provided by CLEO \cite{Anderson:1999wn}, Belle \cite{Abe:2003zm} and FOCUS \cite{Link:2003bd} collaborations, but the separate identification of the two states, together with measurement of their masses and widths,  is due to  Belle  \cite{Abe:2003zm}.
The strange partners,
 first observed in  2003 \cite{Aubert:2003fg}, are very narrow in contrast to expectations.  This  feature  can be attributed to their masses below the $DK$ (for $D_{s0}^*(2317)$) and $D^*K$ (for $D_{s1}^\prime(2460)$) thresholds, so  that the  isospin-conserving  decays are kinematically forbidden. The observed strong decays to $D_s \pi^0$ and $D_s^* \pi^0$ violate the isospin conservation,  hence the narrow widths. Their identification with the doublet  $(D_{s0}^*,\,D_{s1}^\prime)$ is supported by a light-cone QCD sum rule analysis \cite{Colangelo:2005hv} which  reproduces the experimentally observed hierarchy of the radiative decay modes \cite{pdg}.  A puzzling aspect is the mass degeneracy between
the strange states and  their non strange partners \footnote{The mixing between a  $c{\bar s}$  and a four quark configuration could be invoked as an explanation of this degeneracy \cite{Browder:2003fk}. }.
Another issue   is the possible mixing between the two $1^+$ states:  in the case of non strange mesons, the Belle collaboration has determined the mixing angle $\theta$,    with the result:
 $\theta=-0.10 \pm 0.03 \pm 0.02 \pm 0.02 $ rad \cite{Abe:2003zm},  indicating a small mixing.

The other entries in Table \ref{charm} correspond to the most recent observations.
$D_{sJ}(2860)$ was  observed by the BaBar collaboration
\cite{Aubert:2006mh}, and $D_{s1}^{*}(2700)$ by the Belle \cite{Brodzicka:2007aa} and BaBar \cite{Aubert:2006mh} collaborations, both the resonances   in the  $DK$ final state.
The two resonances have been confirmed in $pp$ collisions at the LHC \cite{LHCDsJ}.
The spin-parity $J^P=1^-$ of $D_{s1}^{*}(2700)$ has been established studying the  production in $B$ decays.
$D_{s1}^{*}(2700)$ and $D_{sJ}(2860)$ are  also seen to decay to  $D^*K$ \cite{:2009di}, hence they  have natural parity $J^P=1^-, \, 2^+, \,
3^-, \cdots$.  The  $D^* K$  mode excludes the  assignment $J^P=0^+$ for $D_{sJ}(2860)$.
Additional information comes from the measurements of the ratios of decay rates
\bea
{BR(D_{s1}^{*}(2700) \to D^*K) \over BR(D_{s1}^{*}(2700) \to
DK)} &=& 0.91 \pm 0.13_{stat} \pm 0.12_{syst} \nn \\
{BR(D_{sJ}(2860) \to D^*K) \over BR(D_{sJ}(2860) \to DK)}&=&
 1.10 \pm 0.15_{stat} \pm 0.19_{syst}  \,\, ,
\label{exp1}
\eea
where $D^{(*)}K$ is the sum over the final states $D^{(*)0}K^+$
and $D^{(*)^+}K_S^0$  \cite{:2009di}.  Comparing these data with the
predictions obtained in the heavy quark limit   \cite{Colangelo:2007ds}, we argue that
$D_{s1}^{*}(2700)$ is  the first radial excitation of $D_s^*(2112)$.

The case of $D_{sJ}(2860)$  is more uncertain.
For the possible quantum number assignments to $D_{sJ}(2860)$, one can follow the discussion in \cite{Colangelo:2006rq}. Since this resonance decays to both  $DK$ and $D^*K$,  it may be identified with the lowest lying $n=1$ state with either $J^P_{s_\ell}=1^-_{3/2}$, i.e.  $D_{s1}^*$ in the $X$ doublet, or $J^P_{s_\ell}=3^-_{5/2}$, i.e. the state $D_{s3}$ in the $X^\prime$ doublet.  Another possibility is the identification with the radial excitation with $n=2$ and $J^P_{s_\ell}=2^+_{1/2}$, i.e. the state ${\tilde D}_{s2}^*$ in the ${\tilde T}$ doublet.  Allowed decay modes are into $DK,\,D_s\eta$, $D^*K$ and $D_s^*\eta$.  From  the ratios of strong decay rates in the three possible cases, the identification with $D_{s3}$ was proposed  \cite{Colangelo:2006rq}.
 An important argument concerns the decay width, since $D_{s3}$ is expected to decay in $f$- wave, which would  explain the observed narrow width of the resonance.  $D_{s1}^*$ and ${\tilde D}_{s2}^*$ decay in $p$- and $d$- wave, respectively, which makes  the identification of $D_{sJ}(2860)$ with $D_{s1}^*$ unlikely, but does not exclude that with ${\tilde D}_{s2}^*$. Using typical values for the strong decay constant governing its strong decays, it turns out that indeed $D_{s1}^*$ should have a width  incompatible with the experimental findings.
 As for ${\tilde D}_{s2}^*$, its mass is expected to be larger on the basis of potential model calculations: actually, the prediction $M( {\tilde D}_{s2}^*)\simeq 3.157$ GeV   \cite{Di Pierro:2001uu} must be contrasted to  $M( D_{s3})\simeq 2.925$ GeV.
As a conclusion,   we continue to propose  for $D_{sJ}(2860)$ the  assignment $J^P=3^-$  with  $n=1$.   The alternative  identification with ${\tilde D}_{s2}^*$  is    discussed   in the subsequent Sections.

The BaBar collaboration  observed another  broad structure in the $D^* K$ invariant mass distribution, $D_{sJ}(3040)$,  with too limited statistics to permit
studies of angular distributions for this state  \cite{:2009di}.
Since $D_{sJ}(3040)$  decays to $D^*K$ and it is not found in  the $DK$ distribution,  it has unnatural parity   $J^P=1^+, \, 2^-, \, 3^+, \cdots$.
The lightest    not yet observed states with these quantum numbers  are
the two $J^P=2^-$ states belonging to the $\ell=2$ doublets,   $D_{s2}$ with
$\dd s_\ell^P=\frac{3}{2}^-$ and  $D_{s2}^{\prime *}$ with $\dd s_\ell^P=\frac{5}{2}^-$.  The case
$J^P=3^+$ corresponds to a  doublet with $\dd s_\ell^P=\frac{7}{2}^+$, the mass of which is expected to  be  larger.
In the case of radial excitations, the identification with the states
with $n=2$,  $J^P=1^+$,  and $\dd s_\ell^P=\frac{1}{2}^+$ (the meson ${\tilde
D}_{s1}^\prime$) or   $\dd s_\ell^P=\frac{3}{2}^+$ (the meson ${\tilde D}_{s1}$)  is
possible. Comparison of the features of the four possible classifications for $D_{sJ}(3040)$  shows that, due to the large mass, several decay modes are possible: to a member of the fundamental heavy doublet plus a light pseudoscalar or vector meson and to a member of an excited heavy doublet and a light pseudoscalar meson  \cite{Colangelo:2010te}. In the heavy quark limit,   the two $J^P=1^+$ are expected to be broader than
the two $J^P=2^+$ states,   hence $D_{sJ}(3040)$ is likely to
 be identified with one of the two axial-vector mesons.
A distinction between the two is provided by the  $DK^*$ and $D_s \phi$ decay modes,
since the widths to these final states are  larger for   ${\tilde D}^\prime_{s1}$
than for ${\tilde D}_{s1}$. This justifies the  classification   of $D_{sJ}(3040)$ as one of  the two states with $J^P=1^+$, $n=2$,  proposed in Table \ref{charm}.  The features of the corresponding spin  and non strange partners can be predicted accordingly \cite{Colangelo:2010te}.

The last four states  in  Table  \ref{charm}  are the non-strange $c{\bar q}$ mesons discovered by the BaBar collaboration  in the process  $e^+ e^- \to c{\bar c} \to D^{(*)} \pi X$ \cite{delAmoSanchez:2010vq}.
The four new resonances are found   with   decay modes:
\begin{itemize}
\item
$D^0(2550)$  to $D^{*+} \pi^-$;
\item $D^{*0}(2600)$  to $D^+ \pi^-$ and $D^{*+}\pi^-$,   and the isospin partner $D^{*+}(2600)$  to $D^0 \pi^+$;
\item $D^{*0}(2760)$  to $D^+ \pi^-$,  and the isospin partner $D^{*+}(2760)$  to $D^0 \pi^+$;
\item $D^{*0}(2750)$  to $D^{*+}\pi^-$.
\end{itemize}
Their measured mass and width are reported in Table \ref{charm-pdg}.
The BaBar collaboration has also measured the ratios
\bea
{ {BR}(D^{*0}(2600) \to D^+ \pi^-) \over {BR}(D^{*0}(2600) \to D^{*+} \pi^-)} &=& 0.32 \pm 0.02 \pm 0.09 \label{new4ratios} \\
{ {BR}(D^{*0}(2760) \to D^+ \pi^-) \over {BR}(D^{*0}(2750) \to D^{*+} \pi^-)} &=& 0.42 \pm 0.05 \pm 0.11 \,\,\,\, .
\label{new4ratios2} \eea
In the case of the final state $D^{*+}\pi^-$, important piece of information comes from the distribution in $\cos \theta_H$, with $\theta_H$  the helicity angle between the primary pion $\pi^-$ and the slow pion $\pi^+$ from the $D^{*+}$ decay.
The measured $\cos \theta_H$ distribution for $D^*(2600)$ is consistent with the assignment of natural parity to this state, which also agrees  with the observation in  both $D \pi$ and $D^* \pi$;  moreover,
 the angular distribution for $D^0(2550)$ behaves like $\sim \cos^2 \theta_H$, as expected for a $J^P=0^-$ state.

On the basis of these observations,  the Babar collaboration suggested  that  $(D(2550), \,D^*(2600))$  compose the $\tilde H$,  $J^P=(0^-,1^-)$ doublet of $n=2$ radial excitations of $(D,\,D^*)$ mesons, while $(D(2750),\,D^*(2760))$,  can be identified with  the $\ell=2$, $n=1$ states  \cite{delAmoSanchez:2010vq}, mainly  from comparison of the measured masses with quark model results \cite{Godfrey:1985xj}.  Since there are two possible doublets with $\ell=2$,  the identification with the  $J^P=(2^-,\,3^-)$ doublet would come together with the hypothesis  $D_{sJ}(2860)=D_{s3}$, and in this case $D_{sJ}(2860)$ and $D^*(2760)$   would be  corresponding states with and without strangeness.
We have classified the four new  states in Table \ref{charm} according to these conclusions.
However, we have mentioned that another possibility is  $D_{sJ}(2860)$ being identified with the ${\tilde D}_{s2}^*$ meson. In such a case, if $D^*(2760)$ is  viewed as the non strange partner of $D_{sJ}(2860)$, it can be the state ${\tilde D}_{2}^*$,  and $D(2750)$ its spin partner ${\tilde D}_{1}^\prime$, both filling the doublet ${\tilde T}$.
In the quark model, the masses of the states in the $(D_2^{\prime *},\,D_3)$ doublet are found to be ($2.775$ GeV, $2.799$ GeV), while in the case of the $n=2$ $({\tilde D}_1^\prime\,D_2^*)$ doublet the obtained masses are ($2.995$ GeV, $3.035$ GeV) \cite{Di Pierro:2001uu},  findings that also support  the first one of these two possible assignments.

Let us  turn to the beauty sector. As in the case of charm, the $J^P=(0^-,1^-)$   lightest doublet is well established and is included  in Table \ref{charm} together with the other  observed resonances.   The measured masses and widths of the excited states are reported  in Table \ref{beauty-pdg}.
First observations of open  beauty resonances  were  gained by the LEP collaborations \cite{Akers:1994fz,lep},   using inclusive or semi-exclusive $B$ decays which made  impossible the separation of the states. More recently, the CDF and D0 collaborations at the Tevatron reported evidence of   non strange excited beauty mesons,  which can be identified with the  $(B_1,\,B_2^*)$ components of the $s_\ell^P={3 \over 2}^+$ doublet. They are  found  in  the decays $B^{*0}_2 \to B^+ \pi^-, \, B^{*+} \pi^-$ and $B_1^0 \to B^{*+}\pi^-$ \cite{Abazov:2007vq,:2008jn}.
Analogously,  for strange-beauty mesons the  first  observation of $p$-wave states was not able to separate the individual components \cite{Akers:1994fz}. Recent  CDF and D0 studies have reported  evidence of $B_{s1}$ decaying to $B^{*+}K^-$,  and of $B_{s2}^*$ decaying to $B^+ K^-$  \cite{:2007tr,:2007sna}.
These resonances  can be assigned  to the $ s_\ell^P=\frac{3}{2}^+$ beauty doublet with  strangeness. Confirmation of these orbitally excited $B$ and $B_s$ mesons, with compatible masses and widths, has  been obtained by the LHCb collaboration \cite{Pappagallo-LHCb}.
\begin{table*}[h!]
\centering \caption{Measured mass and width of the observed open beauty  excited  mesons \cite{pdg}.}\label{beauty-pdg}
\begin{tabular}{|c  c c | c c c | }\hline
$b{\bar q}$ &   mass (MeV) &  $\Gamma$ (MeV) & $b{\bar s}$  &  mass (MeV)\hskip 1 cm  &  $\Gamma$ (MeV)  \\ \hline
$\,\,\,\, B_1^0(5721)\,\,\,\, $ & $\,\,\,\, 5723.4 \pm2.0\,\,\,\, $ &  & $\,\,\,\, B_{s1}^0(5830)\,\,\,\, $ & $\,\,\,\, 5829.4 \pm 0.7\,\,\,\, $& \\\hline
$B_2^{*0}(5747)$ & $5743 \pm 5$ & $ \,\,\,\, 22.7 \pm^{3.8}_{3.2} \pm^{3.2}_{10.2}\,\,\,\, $ & $B_{s2}^{*0}(5840)$  & $5839.7 \pm 0.6$ & \\ \hline
   \end{tabular}
\end{table*}

\section{Meson mass parameters and predictions}\label{section:masses}

The HQ symmetries and the mass measurements  permit to predict  the masses of not yet observed  open charm and open beauty resonances,  filling   a few empty spaces in Table \ref{charm}.
The preliminary  step is to determine  the average masses ${\bar M}_F$ in Eqs.(\ref{masse-medie}), the mass splittings $\Delta_F$
between doublets and the splittings $\lambda_F$ between spin partners in a  doublet  defined in Eqs.(\ref{lambdat})  for all the observed states in Table \ref{charm}.
The results are collected in Table \ref{splittings}.  Flavour symmetry implies that the mass splittings $\Delta_F$
between doublets  are the same regardless of the heavy quark flavour. Furthermore,  also  the mass splitting  $\lambda_F$ between spin partners  in a  doublet should not depend on the
heavy flavour. Considering the various entries in  Table  \ref{splittings}, we see  that these statements are experimentally violated  both by light flavour and  heavy quark mass effects, since such entries incorporate  higher order  symmetry breaking terms: in particular, the strange quark mass effect is  clearly visible in the average mass parameters ${\bar M}_F$.

Using this input, we can elaborate several predictions for the masses of unobserved states.
Let us consider ${\tilde D}_s$, the spin partner of ${\tilde D}^{*}_s$ which we have identified with $D_{s1}^{*}(2700)$.  If the effect of the strange quark is to shift the mass of a given state by the same amount in the fundamental  and in the  $n=2$ radial excitation doublet, we have $M_{D_s}-M_{D^0}=M_{{\tilde D}_s}-M_{{\tilde D}^{ 0}}$.  Identifying ${\tilde D}^{0}$ with the observed $D^0(2550)$, we can infer that
$M_{{\tilde D}_s}=2643 \pm 8 \,\, {\rm MeV}$.
The consistency of the identification of $D^{*0}(2600)$ with the spin partner of $D^0(2550)$, i.e. with the state ${\tilde D}^{* 0}$,  can analogously be checked. From  $M_{D^*_s}-M_{D^{*0}}=M_{{\tilde D}^{*}_s}-M_{{\tilde D}^{* 0}}$ we obtain: $M_{{\tilde D}^{* 0}}=2604 \pm 9$ MeV,
which supports the identification of  $D^{*0}(2600)$ with ${\tilde D}^{* 0}$.  A  compatible result is obtained using the $\lambda_{\tilde H}$ parameter  in Table  \ref{splittings} to predict the ${\tilde D}_s$ mass:
$M_{{\tilde D}_s}=2643 \pm 13$ MeV.  This allows us to conclude that
the masses of all   $n=1$ and $2$,  $J^P_{s_\ell}=(0^-,\,1^-)_{1/2}$ charmed mesons  with and without strangeness,   are determined, see Table \ref{cmasses}.

In the case of the doublet  $J^P_{s_\ell}=(2^-,\,3^-)_{5/2}$, if the two states $D(2750)$ and $D^*(2760)$  are the  $(D_2^{\prime *},\,D_3)$ members of this $\rm X^\prime$,   $c{\bar q}$ doublet, we  can predict the mass of the spin partner $D_{s2}^{\prime *}$ of $D_{sJ}(2860)$ identified with $D_{s3}$.   From
$M_{D_{s2}^{\prime *}}=M_{D_{s3}}-(M_{D_{3}}-M_{D_2^{\prime *}})$ we have:
$M_{D_{s2}^{\prime *}}=2851 \pm 7 \,\, {\rm MeV}$.
 This prediction for the mass of the spin partner of $D_{sJ}(2860)$ holds independently of the identification of the latter, it  only  relies on the assumption that $D_{sJ}(2860)$ and $D^*(2760)$ have the same quantum numbers and differ  for the strangeness.

\begin{table*}[b]
\centering \caption{Values of the spin averaged masses ${\bar M}_F$ (in MeV), of the mass splittings $\Delta_F$ (in MeV) and of the hyperfine splitting parameters  $\lambda_{F}$ (in MeV$^2$) defined in Eqs.(\ref{masse-medie}),(\ref{lambdat}). }
    \label{splittings}
   \begin{tabular}{|lcccccc|}
      \hline
  & $c{\bar u}$  &$c{\bar d}$  & $c{\bar s}$ &$b{\bar u}$ & $b{\bar d}$ & $b{\bar s}$\\ \hline
  $\overline M_{H}$ & $1971.45\pm 0.12 \,$ & $1975.12\pm 0.10 \,$ & $2076.4\pm 0.4 \,$ & $5313.7\pm  0.3\,$ & $5313.8\pm  0.3\,$ & $5403\pm 2 \,$ \\
  $\overline M_{\tilde H}$ & $2591.4\pm 3.3 \,$ &  &   &  &  &  \\
$\overline M_{S}$ & $2400\pm 28 \,$ &  & $2424.1\pm 0.5 \,$ & & & \\
$\overline M_{T}$ & $2447.1\pm 0.5 \,$ & $ 2449.0 \pm 1.6  \,$ &$2558.6 \pm 0.6 \,$ & &$5735.7 \pm 3.2 \,$ &$5834.7 \pm 0.5 \,$ \\
$\overline M_{X^\prime}$  & $2758.8\pm 2.3 \,$ & & & & & \\
\hline
$ \Delta_{S}$ & $429\pm 28 \,$ &  &$347.7\pm 0.6 \,$ & & &\\
$ \Delta_{T}$ & $475.7\pm 0.5 \,$ &$473.9\pm 1.6 \,$ & $482.2\pm 0.7 \,$ & & $421.9\pm 3.2 \,$&$431.7\pm 2.1 \,$\\
 $ \Delta_{X^\prime}$& $787.4\pm 2.3 \,$ &  &  & && \\
 \hline
   $\lambda_H$ & $\,\,\,\, (262.3 \pm 0.2)^2\,\,\,\, $ & $\,\,\,\, (261.2 \pm 0.2)^2\,\,\,\, $ &
 $\,\,\,\, (270.9 \pm 0.6)^2\,\,\,\, $
 & $\,\,\,\, (246.8 \pm 1.2)^2\,\,\,\, $ & $\,\,\,\, (245.9 \pm 1.2)^2\,\,\,\, $ & $\,\,\,\, (256.3 \pm 6.4)^2\,\,\,\, $\\
 $\lambda_{\tilde H}$ & $ (211.2 \pm \,\,13.4)^2$ &
 &  &  & & \\
$\lambda_S$ & $ (254 \pm 54)^2$ &  & $(290.9 \pm 0.9)^2$
 &  & &\\
$\lambda_T$ & $(195\pm 2)^2$ & $ (193 \pm 7)^2$&$ (189.2 \pm 2.1)^2$ & &$(205\pm 28)^2$ &$(149.9\pm 6.7)^2$\\
$\lambda_{X^\prime}$  & $\,\,\,\, (112\pm 24)^2\,\,\,\, $ && & &  & \\   \hline
  \end{tabular}
\end{table*}
Using the estimated masses of ${\tilde D}_s$ and $D_{s2}^{\prime *}$  we obtain: ${\bar M}_{\tilde H}=2692.5\pm 7.0 \, {\rm  MeV}$, $\Delta_{\tilde H}=616 \pm 7$ MeV, $\lambda_{\tilde H}= (210 \pm 19 \,\, {\rm MeV})^2$,  and ${\bar M}_{X^\prime}= 2857.5\pm 4.3 \, {\rm MeV}$, $\Delta_{X^\prime}=781 \pm 4$ MeV, $\lambda_{X^\prime}=(114 \pm 46 \,\, {\rm MeV})^2$ for $c{\bar s}$ mesons.
Another possibility is that $D_{sJ}(2860)$ and  $(D(2750),\,D^*(2760))$ belong to the $n=2$,  $\tilde T$ doublets: In this case  we  would get
$\overline M_{\tilde T}=2759.2\pm 2.4 \, {\rm MeV}$,
$\Delta_{\tilde T}=787.8\pm 2.4 \, {\rm MeV}$ and  $\lambda_{\tilde T} = (106 \pm 22 \,\, {\rm MeV})^2$.
\begin{table*}[h!]
\centering  \caption{Predicted mass  and width  (in MeV) of two not yet observed excited charm mesons, quoted together with the other members of their respective  doublets.}
    \label{cmasses}
   \begin{tabular}
{| c c |cc| cc |}  \hline
 & &${\tilde D}_{(s)} \,\, (0^-,n=2)$ & ${\tilde D}_{(s)}^{*}\,\, (1^-,n=2)$&  $D_{(s)2}^{\prime *}\,\, (2^-)$& $D_{(s)3}\,\, (3^-)$ \\ \hline
 $c \bar q$\,\,& & $D(2550)$ & $D^*(2600)$&  $D(2750)$ & $D(2760)$\,\, \\
 $c \bar s$\,\,& mass&$ 2643 \pm 13$\,\, \,& $D^*_{s1}(2700)$& $ 2851 \pm 7$\,\,\, & $D_{sJ}(2860)$ \\
  &$\Gamma$ &$ 33.5 \pm 3.3$\,\, \,& & $ 20.5 \pm 2.4$\,\,\, &  \\
\hline
 \end{tabular}
\end{table*}

In the HQ limit,   charm data can be exploited  to make predictions for beauty.  The procedure we adopt, based on the Lagrangian (\ref{L}),(\ref{mass-viol}),  is to  assume the equalities
\bea
\Delta_F^{(c)}&=&\Delta_F^{(b)} \nn\\
\lambda_F^{(c)}&=&\lambda_F^{(b)} \nn
\eea
for $F={\tilde H}, \,S,\,T$, $X^\prime$ and ${\tilde T}$, and to use these two expressions, with the l.h.s. experimentally determined, to predict the masses of the two states in the corresponding beauty doublets.
The results  are collected in Table \ref{bmasses}.
If $D_{sJ}(2860)$ and  $(D(2750),\,D^*(2760))$) are assigned to the $n=2$, $\tilde T$  doublets,  the last two columns in this Table would represent  the predictions for  ${\tilde B}_{(s)1}^\prime$ and ${\tilde B}_{(s)2}$, respectively.
It is worth remarking that the masses of $B_{s0}^*$ and $B_{s1}^\prime$   are below the $B K$ and $B^{*} K$ thresholds,
and therefore these two mesons  are expected to  be  very narrow, with main decays into   $B_{s} \pi^0$ and $B_{s}^* \pi^0$   \cite{Colangelo:2003vg,Colangelo:2004vu,Colangelo:2005gb}.

\begin{table*}[h!]
\centering  \caption{Predicted mass  and width (in MeV) of doublets of excited beauty mesons. For the decay widths of $B^*_{s0}$ and $B^\prime_{s1}$ see the text.}
    \label{bmasses}
   \begin{tabular}
{|cc|cc|cc|cc|}
      \hline
 &&${\tilde B}_{(s)} \,\, (0^-,n=2)$ & ${\tilde B}_{(s)}^{*}\,\, (1^-,n=2)$& $ B^*_{(s)0} \,\, (0^+)$ & $ B^\prime_{(s)1}  \,\, (1^+)$  & $B_{(s)2}^{\prime *}\,\, (2^-)$& $B_{(s)3}\,\, (3^-)$ \\ \hline
 $b \bar q$\,\,&mass&$5911.1 \pm 4.9$\,\,\, & $5941.2 \pm 3.2$ \,\,\,& $5708.2\pm 22.5 $ \,\,&$5753.3\pm 31.1 $\,\,\, &  $ 6098.2\pm2.4$\,\,\, & $6103.1\pm2.6 $\,\,\, \\
 &$\Gamma$&$149 \pm 15$\,\,\, & $186\pm18$ \,\,\,& $269\pm58 $ \,\,&$268\pm 70\ $\,\,\, &  $ 103\pm8$\,\,\, & $129\pm10 $\,\,\, \\
 &         &                                 &                               &                                  &                                   &                                          &               \\
 $b \bar s$\,\,& mass&$ 5997.3 \pm 6.1$\,\, \,& $6026.6 \pm 7.9 $\,\,\,& $5706.6\pm 1.2$\,\, \,&$5765.6\pm 1.2$\,\,\, &$ 6181.3\pm5.2$\,\,\, & $6186.3\pm4.6$\,\,\, \\
 &$\Gamma$&$ 76 \pm 9$\,\, \,& $118 \pm 14 $\,\,\,& \,\, \,&\,\,\, & $57\pm6$\,\,\, & $78.4\pm7.3$\,\,\, \\
\hline
    \end{tabular}
\end{table*}

\section{Strong two-body Decays to $\rm H$  and a light pseudoscalar meson}\label{c-decays}
To describe  the decays  $F \to H M$, with $F=H,S,T,X,X^\prime$
and $M$  a light pseudoscalar meson,  at the leading order in the  light meson
momentum and heavy quark mass expansion,  we can employ the Lagrangian interaction  terms \cite{hqet_chir}:
\bea
{\CMcal L}_H &=& \,  g \, Tr \Big[{\bar H}_a H_b \gamma_\mu\gamma_5 {\CMcal A}_{ba}^\mu \Big] \nn \\
{\CMcal L}_S &=& \,  h \, Tr \Big[{\bar H}_a S_b \gamma_\mu \gamma_5 {\CMcal A}_{ba}^\mu \Big]\, + \, h.c.  \nn \\
{\CMcal L}_T &=&  {h^\prime \over \Lambda_\chi}Tr\Big[{\bar H}_a T^\mu_b (i D_\mu {\spur {\CMcal A}}+i{\spur D} { \CMcal A}_\mu)_{ba} \gamma_5\Big] + h.c.    \nn \\
{\CMcal L}_X &=&  {k^\prime \over \Lambda_\chi}Tr\Big[{\bar H}_a X^\mu_b (i D_\mu {\spur {\CMcal A}}+i{\spur D} { \CMcal A}_\mu)_{ba} \gamma_5\Big] + h.c.   \,\,\,\,\,\,\,\,\,\,\,\,   \label{lag-hprimo} \\
{\CMcal L}_{X^\prime} &=&  {1 \over {\Lambda_\chi}^2}Tr\Big[{\bar H}_a X^{\prime \mu \nu}_b \big[k_1 \{D_\mu, D_\nu\} {\CMcal A}_\lambda + k_2 (D_\mu D_\lambda { \CMcal A}_\nu + D_\nu
D_\lambda { \CMcal A}_\mu)\big]_{ba}  \gamma^\lambda \gamma_5\Big] + h.c.  \nn
 \eea
 The chiral symmetry-breaking scale $\Lambda_\chi$ is set to
$\Lambda_\chi = 1 \, $ GeV. ${\CMcal L}_S$ and ${\CMcal L}_T$ describe
transitions of positive parity heavy mesons with the emission of
light pseudoscalar mesons in $s-$ and $d-$ wave, respectively,
and $g, h$ and $h^\prime$ are    effective coupling constants.
${\CMcal L}_X$ and ${\CMcal L}_{X^\prime}$
describe the transitions of higher mass mesons of negative parity, belonging to the $ X$ and $ X^\prime$ doublets,
with the emission of light pseudoscalar mesons in $p-$ and $f-$
wave,  with coupling constants $k^\prime$, $k_1$ and $k_2$ (we set $k=k_1+k_2$). At the same  order in the expansion in the light meson momentum,
the structure of the Lagrangian terms for radial excitations of
the various doublets does not change since it is only dictated by the spin-flavour and chiral symmetries, but  the
coupling constants  must be  replaced by new ones  denoted by $\tilde g$, $\tilde h$, and so on.
This formulation is useful since meson transitions  into final states
obtained  by flavour  and heavy quark spin trasnformations can be related
in a straightforward way. The expressions of the decay widths obtained  from  Eqs.(\ref{lag-hprimo}),  considering the various doublets which the decaying meson belongs to, are the following:
\begin{itemize}
\item Decaying meson   $H=(P,P^*)$ or ${\tilde H}=({\tilde P},{\tilde P}^{ *})$:
\bea
\Gamma(P^* \to P M) &=& C_M { g^2 \over 6 \pi f_\pi^2}{M_P \over M_{P^*}} |{\vec p}_M|^3 \label{1-Hto0-0-} \\
\Gamma(\tilde P^* \to P M) &=& C_M { \tilde g^2 \over 6 \pi f_\pi^2}{M_P \over M_{\tilde P^*}} |{\vec p}_M|^3
\label{1-Hto0-0-tilde}
\\
\Gamma(\tilde P^*_i \to P^*_f M) &=& C_M {\tilde g^2 \over 3 \pi f_\pi^2}{M_{P^*_f} \over M_{\tilde P^*_i}} |{\vec p}_M|^3        \label{1-Hto1-0-}
\\
\Gamma(\tilde P \to P^* M) &=& C_M {\tilde g^2 \over 2 \pi f_\pi^2} {M_{P^*} \over M_{\tilde P}} |{\vec p}_M|^3 \,\,\,\, .\label{0-Hto1-0-}
\eea
\item Decaying  $S=(P^*_0,P^\prime_1)$:
\bea
\Gamma(P^*_0 \to P M) &=& C_M { h^2 \over2 \pi f_\pi^2}{M_P \over M_{P^*_0}} \left[m_M^2+|{\vec p}_M|^2 \right] \,|{\vec p}_M| \label{0+Sto0-0-}
\\
\Gamma(P^\prime_1 \to P^* M) &=& C_M { h^2 \over2 \pi f_\pi^2}{M_{P^*} \over M_{P^\prime_1}} \left[m_M^2+|{\vec p}_M|^2 \right] \,|{\vec p}_M| \,\,\,\, .\label{1+Sto1-0-}
\eea
\item Decaying   $T=(P_1,P^*_2)$ or  ${\tilde T}=({\tilde P}_1,{\tilde P}^*_2)$:
\bea
\Gamma(P_1 \to P^* M) &=& C_M { 2 h^{\prime 2} \over 3 \pi f_\pi^2}{M_{P^*} \over M_{P_1}} |{\vec p}_M|^5 \label{1+Tto1-0-}
\\
\Gamma(P_2^* \to P M) &=& C_M { 4 h^{\prime 2} \over 15 \pi f_\pi^2}{M_{P} \over M_{P_2^*}} |{\vec p}_M|^5 \label{2+Tto0-0-}
\\
\Gamma(P_2^* \to P^* M) &=& C_M { 2 h^{\prime 2} \over 5 \pi f_\pi^2}{M_{P^*} \over M_{P_2^*}} |{\vec p}_M|^5 \,\,\,\, .\label{2+Tto1-0-}
\eea
\item Decaying   $X=(P_1^*,P_2)$:
\bea
\Gamma(P^*_1 \to P M) &=& C_M { 4 k^{\prime 2 } \over 9 \pi f_\pi^2}{M_P \over M_{P^*_1}} \left[m_M^2+|{\vec p}_M|^2 \right] \,|{\vec p}_M|^3 \label{1-Xto0-0-}
\\
\Gamma(P^*_1 \to P^* M) &=& C_M { 2 k^{\prime 2 }  \over 9 \pi f_\pi^2}{M_{P^*} \over M_{P^*_1}} \left[m_M^2+|{\vec p}_M|^2 \right] \,|{\vec p}_M|^3 \label{1-Xto1-0-}
\\
\Gamma(P_2 \to P^* M) &=& C_M { 2 k^{\prime 2 }  \over 3 \pi f_\pi^2}{M_{P^*} \over M_{P_2}} \left[m_M^2+|{\vec p}_M|^2 \right] \,|{\vec p}_M|^3 \,\,\,\, .\label{2-Xto1-0-}
\eea
\item Decaying    $X^\prime=(P_2^{\prime *},P_3)$:
\bea
\Gamma(P_2^{\prime *} \to P^* M) &=& C_M { 4 k^2 \over 15 \pi f_\pi^2}{M_{P^*} \over M_{P_2^{\prime *}}} |{\vec p}_M|^7 \label{2-Xpto1-0-}
\\
\Gamma(P_3 \to P M) &=& C_M { 4 k^2 \over 35 \pi f_\pi^2}{M_{P} \over M_{P_3}} |{\vec p}_M|^7 \label{3-Xpto0-0-}
\\
\Gamma(P_3 \to P^* M) &=& C_M { 16 k^2 \over 105 \pi f_\pi^2}{M_{P^*} \over M_{P_3}} |{\vec p}_M|^7 \,\,\,\, .
\label{3-Xpto1-0-}
\eea
\end{itemize}
The coefficients $C_M$ are different for the various light pseudoscalar mesons: $C_{\pi^+}=C_{K^+}=1$, $C_{\pi^0}=C_{K_S}={1 \over 2}$, $C_{\eta}={2 \over 3}$.  ${\vec p}_M$ is the three momentum of $M$.
Notice that only for the states ${\tilde P}^{ *}$ in ${\tilde H}$, $P^*_2$ in $T$, $P^*_1$ in $X$ and $P_3$ in $X^\prime$, both the decays to $P\,M$ and $P^*\,M$ are allowed, while  the other resonances can decay either to $P\,M$ or to $P^*\,M$, a useful observation for the classification of the  states.

The decay rates  depend on effective coupling constants which need to be specified.   Model dependence  in a determination of such couplings can be avoided considering ratios of widths in which  the constants cancel out. Therefore,
for each  meson $F_{(s)}$ that can decay both to $P_{(s)}\,M$ and $P^*_{(s)}\,M$, we focus on the following ratios, considering as reference modes the decay to $D\pi$ for the non strange mesons, and to $DK$ for the strange ones:
\bea
R_\pi^{(F)}&=&\displaystyle{BR(F \to D^* \pi) \over BR(F \to D \pi)}  \,\,\,\, , \nn \\
R_K^{(F_s)}&=&\displaystyle{BR(F_s \to D^* K) \over BR(F_s \to D K)}  \,\,\, ,  \hskip 1.5 cm
R_\eta^{(F_s)}=\displaystyle{BR(F_s \to D_s \eta) \over BR(F_s \to D K)}  \,\,\, ,  \hskip 1.5 cm
R_\eta^{*(F_s)}=\displaystyle{BR(F_s \to D_s^* \eta) \over BR(F_s \to D K)}  \,\,\, .
\label{R-pi-K-eta}
 \,\, \eea
 $D^{(*)}\pi(K)$ indicates $D^{(*)0}\pi^+(K^+)+D^{(*)+}\pi^0(K_S)$ for charged states, and $D^{(*)0}\pi^0(K_S)+D^{(*)+}\pi^-(K^-)$ for neutral
 ones.  Such  ratios can be experimentally determined, so a comparison with the theoretical outcome is possible.

 In Table \ref{ratios-charm}  we collect the predictions for the charmed states $D^*(2600)$ and $D_{s1}^*(2700)$, identified with ${\tilde D}^{ *}$ and ${\tilde D}^{ *}_s$, respectively; for  $D_2^{*0}(2460)$ and $D_{s2}^*(2573)$  and  for $D^*(2760)$ and $D_{sJ}(2860)$ identified with the states $D_3$ and  $D_{s3}$, as assumed in Table \ref{charm}.
 The case of $D^*_{(s)1}$ in the doublet $X$ is not included in the Table since there are at present no  candidates for it;  however, we consider it in the following.
 \begin{table*}[h!]
\centering  \caption{Theoretical ratios $R_M^{(F)}$ for charmed and beauty mesons. The results  are obtained identifying $D^{*0}(2760)$ as  $D_3$ and $D_{sJ}(2860)$ as $D_{s3}$.}
    \label{ratios-charm}
   \begin{tabular}
{|c|c||c|ccc|} \hline
$c{\bar q}$  & \,\,$R_{\pi}$\,\, & $c{\bar s}$  & \,\,$R_{K^0} $\,\, &\,\, $R_\eta$\,\, &\,\, $R_\eta^*$\,\,\\
      \hline
     \,\, $D^{*0}(2600)$\,\, & \,\,$1.22 \pm 0.01$\,\, & \,\,$D_{s1}^{*}(2700)$\,\, & \,\, $0.91 \pm 0.03$\,\,&  \,\, $ 0.195 \pm 0.006$   \,\,& \,\, $ 0.05 \pm 0.01$  \,\,\\
     \,\, $D^{*0}_2(2460)$ \,\,& \,\,$0.440 \pm 0.001$ \,\,&\,\, $D_{s2}^*(2573)$\,\, &\,\,$0.086 \pm 0.002$\,\, & \,\, $0.018 \pm 0.001$\,\, & -\\
    \,\,  $D^{*0}(2760)$\,\,&\,\, $0.514 \pm 0.004$\,\, & \,\,$D_{sJ}(2860)$\,\, & \,\, $ 0.39 \pm 0.01$\,\,& \,\, $ 0.132 \pm 0.003$\,\, & \,\, $0.025 \pm 0.001$ \,\,\,\\
\hline \hline
$b{\bar q}$  & \,\,$R_{\pi}$\,\, & $b{\bar s}$  & \,\,$R_{K} $\,\, &\,\, $R_\eta$\,\, &\,\, $R_\eta^*$\,\,\\
      \hline
     \,\, ${\tilde B}^{*}$\,\, & \,\,$1.63 \pm 0.005$\,\, & \,\,${\tilde B}_s^{*}$\,\, & \,\, $1.43 \pm 0.015$\,\,&  \,\, $ 0.132 \pm 0.008$   \,\,& \,\, $ 0.11 \pm 0.015$  \,\,\\
     \,\, $B^{*}_2$ \,\,& \,\,$0.87 \pm 0.01$ \,\,&\,\, $B_{s2}^*$\,\, &\,\,$0.07 \pm 0.005$\,\, & \,\, -\,\, & \,\,- \,\,\\
    \,\,  $B_3$ \,\,&\,\, $0.92 \pm 0.005$\,\, & \,\,$B_{s3}$\,\, & \,\, $ 0.815 \pm 0.006$\,\,& \,\, $ 0.103 \pm 0.002$\,\, & \,\, $0.063 \pm 0.003$ \,\,\\
\hline
    \end{tabular}
\end{table*}
In the alternative classification  of
 $D^{*0}(2760)$ as ${\tilde D}_2^{*}$ we  obtain $R_\pi=0.775 \pm 0.003$;  analogously, identifying $D_{sJ}(2860)$ with ${\tilde D}_{s2}^{*}$ we get  $R_K= 0.63 \pm 0.01$, $R_\eta= 0.19 \pm 0.002$ and $R_\eta^*=0.07 \pm 0.003$.

For some of the resonances considered in Table \ref{ratios-charm} experimental data are available.
For  $D^*_2(2460)$ one has  \cite{pdg}:
\bea
{\Gamma_1 \over \Gamma_2}&=&{\Gamma(D^*_2(2460)^\pm \to D^0 \pi^+) \over \Gamma(D^*_2(2460)^\pm \to D^{*0} \pi^+) }=1.9 \pm 1.1 \pm 0.3 \nn \\
R_{12}&=&{\Gamma_1 \over \Gamma_1 + \Gamma_2}=0.62  \pm 0.03 \pm 0.02  \eea
for the charged state,  and
\bea
{\Gamma_1 \over \Gamma_2}&=&{\Gamma(D^*_2(2460)^0 \to D^+ \pi^-) \over \Gamma(D^*_2(2460)^0 \to D^{*+} \pi^-) }=1.56 \pm 0.16 \pm 0.3 \nn \\
R_{12}&=&{\Gamma_1 \over \Gamma_1 + \Gamma_2}=0.62 \pm 0.03 \pm 0.02  \eea
for the neutral one.  For these quantities the theoretical results are
\bea
\left({\Gamma_1 \over \Gamma_2}\right)_{charged}&=& 2.266 \pm 0.015 \hskip 1cm (R_{12})_{charged} = 0.694 \pm 0.001 \nn \\
 \left({\Gamma_1 \over \Gamma_2}\right)_{neutral}&=&2.280 \pm 0.007 \hskip 1cm (R_{12})_{neutral} = 0.695 \pm 0.001 \,\,.
\eea
The comparison of  predictions with data  shows a deviation from the heavy quark limit  in the case of the neutral channel. However, the  PDG result  is obtained averaging several data in the range $\left(\displaystyle{\Gamma_1 \over \Gamma_2}\right)_{neutral} \in [2.2-3.0]$, and the average is dominated by a single measurement provided by BaBar collaboration \cite{delAmoSanchez:2010vq},  which requires a  confirmation.

The corresponding charmed-strange resonance is  $D_{s2}^*(2573)$, identified as the $2^+$ member of the $\ell=1$, $s_\ell^P=\displaystyle{3 \over 2}^+$,  $T$ doublet.
It decays to $D^{(*)+}K_S$, $D^{(*)0}K^+$, $D_s \eta$, and is below the $ D_s^* \eta$  threshold.
There are no experimental data for the ratios in Table \ref{ratios-charm}; however, the PDG quotes the upper bound:
$\displaystyle{{BR}(D^{*}_{s2}(2573) \to D^{*0} K^+) \over {BR}(D^{*}_{s2}(2573) \to D^0 K^+)}<0.33$,  which is satisfied by our result
$\displaystyle {{BR}(D^{*}_{s2}(2573) \to D^{*0} K^+) \over {BR}(D^{*}_{s2}(2573) \to D^0 K^+)}=0.091 \pm 0.002  \,\,.$

The  $D_{s1}^{*}(2700)$ has been  treated in Ref.\cite{Colangelo:2007ds}, where the ratios in Table \ref{ratios-charm}  were analyzed.
The BaBar measurement of the first ratio in Eq.(\ref{exp1})   \cite{:2009di} is in agreement with the theoretical outcome reported in the Table,  and this supports  the classification of this state as ${\tilde D}_{s1}^{*}$.

\begin{figure}[t]
 \begin{center}
  \includegraphics[width=0.31\textwidth] {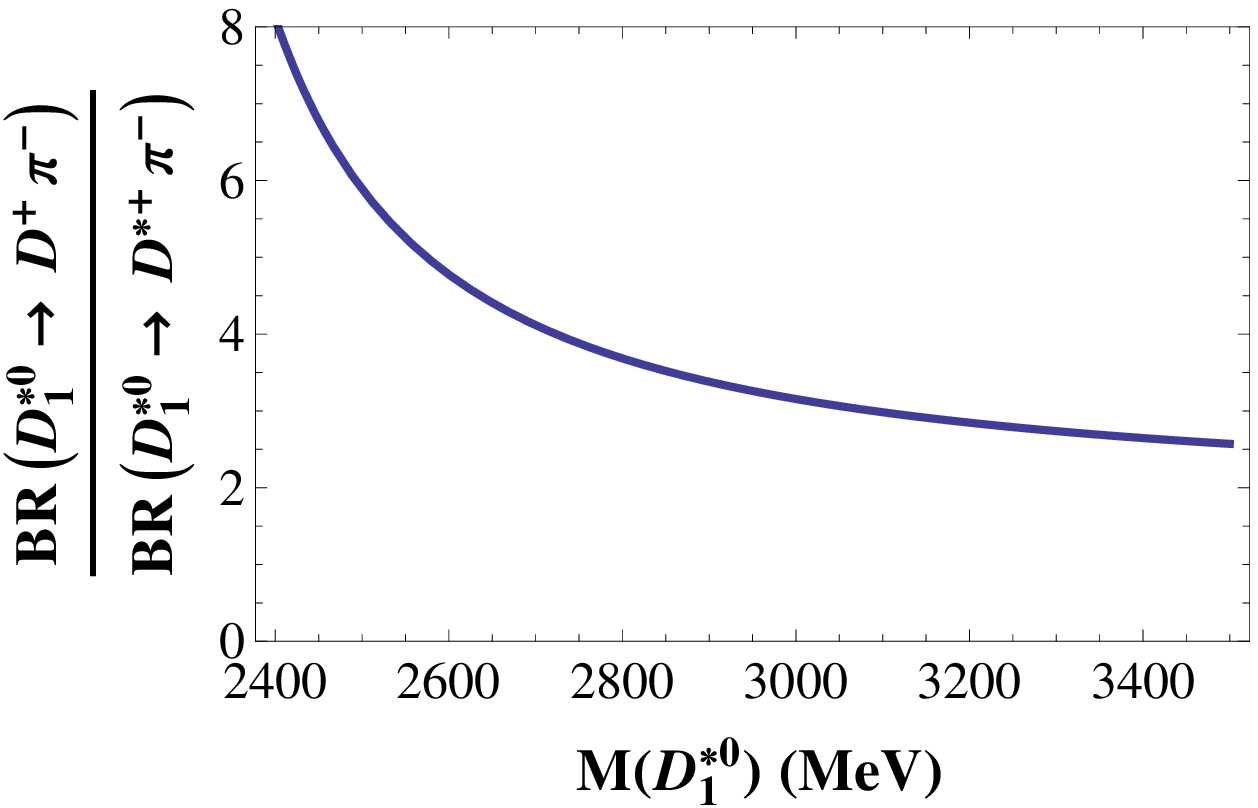}\hspace*{0.3cm}
  \includegraphics[width=0.31\textwidth] {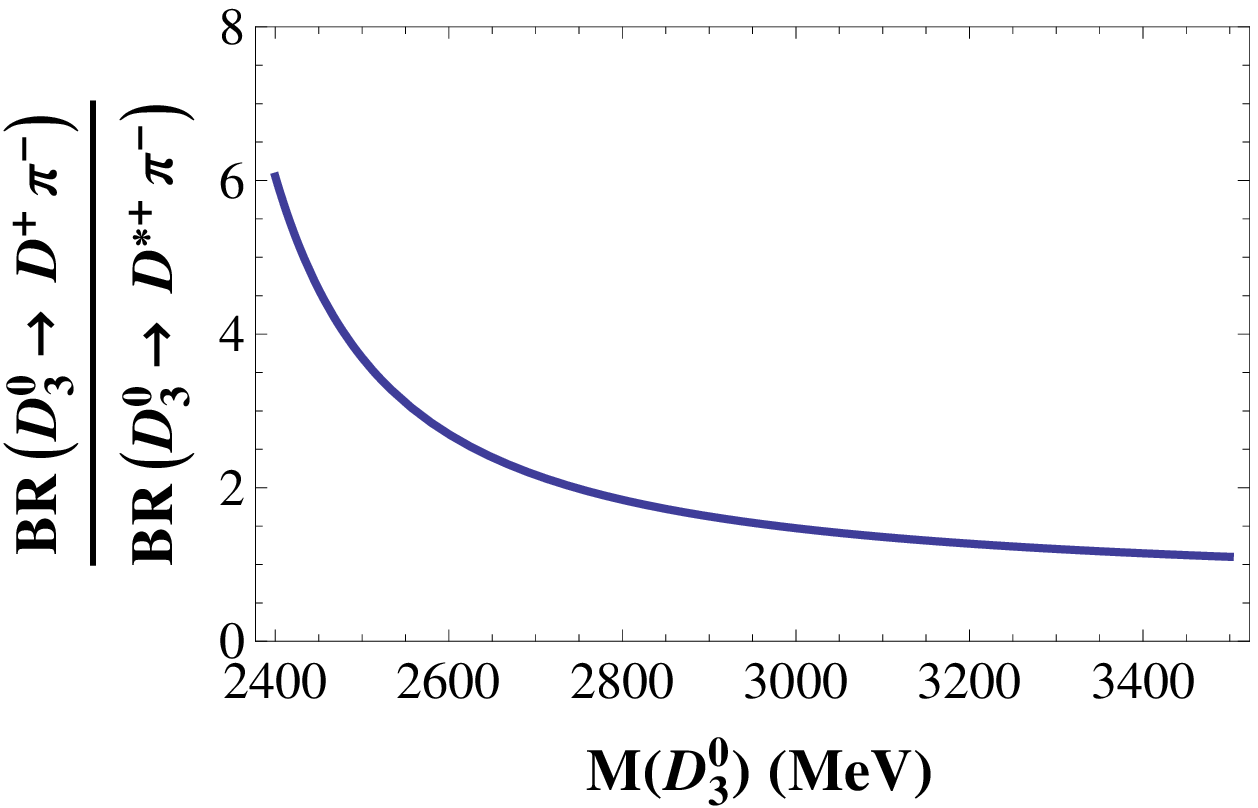} \hspace*{0.3cm}
   \includegraphics[width=0.31\textwidth] {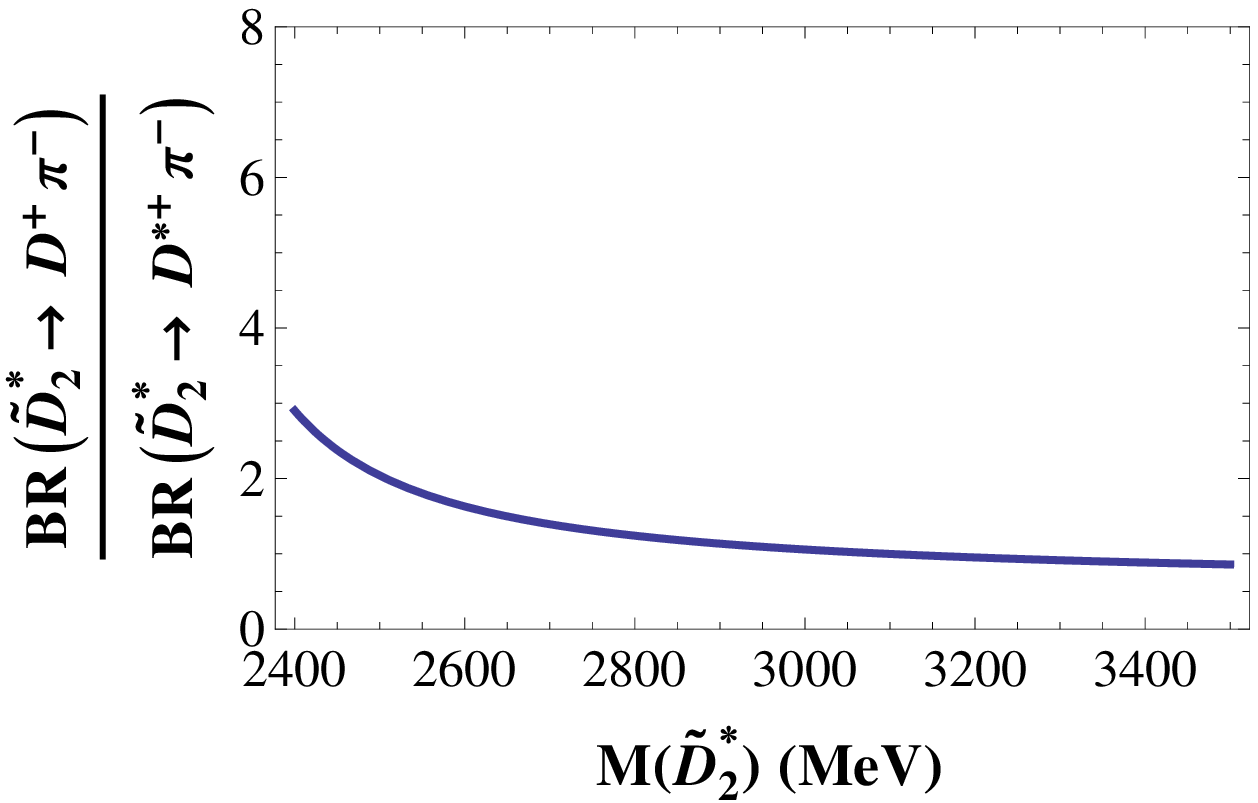}
\vspace*{0mm}
 \caption{Ratios $\displaystyle{ {BR}(D^{*}_1 \to D^+ \pi^-) \over { BR}(D^{*}_1 \to D^{*+} \pi^-)}$, $\displaystyle{ {BR}(D_3 \to D^+ \pi^-) \over {BR}(D_3 \to D^{*+} \pi^-)}$ and $\displaystyle{ {BR}({\tilde D}_2^{*0} \to D^+ \pi^-) \over {BR}({\tilde D}_2^{*0} \to D^{*+} \pi^-)}$ versus the mass of the decaying meson. }
  \label{RXXp}
 \end{center}
\end{figure}

Let us turn to  the first radial excitation of $D^*$, the  ${\tilde D}^{*}$. As discussed in the previous Sections, there are hints that  the observed $D^*(2600)$ might be identified with such a state, and in this case
 the theoretical  ratio  Eq.(\ref{new4ratios}) (identifying $D^*(2600)^0$ with ${\tilde D}^{* 0}$) would be
\be
{ {BR}(D^{*0}(2600) \to D^+ \pi^-) \over {BR}(D^{*0}(2600) \to D^{*+} \pi^-)}=0.822 \pm 0.003 \label{R2600} \,\,\, , \ee
which is  larger than the measurement in (\ref{new4ratios}). To investigate the origin of the discrepancy, we consider   the other   states (up to $\ell=2$)   for which the decays to $P\,M$ and $P^*\,M$ are both allowed, namely
$P^*_1$ in $X$, $P_3$ in $X^\prime$ and ${\tilde P}_2^*$ in $\tilde T$.
In Fig.\ref{RXXp} we plot  the ratios $\displaystyle{ {BR}(D^{*0}_1 \to D^+ \pi^-) \over {BR}(D^{*0}_1 \to D^{*+} \pi^-)}$,  $\displaystyle{ {BR}(D_3^0 \to D^+ \pi^-) \over {BR}(D_3^0 \to D^{*+} \pi^-)}$ and $\displaystyle{ {BR}({\tilde D}_2^{*0} \to D^+ \pi^-) \over {BR}({\tilde D}_2^{*0} \to D^{*+} \pi^-)}$ versus the mass of the decaying meson.
The  ratio  exceeds  $1$ both for $D_3$ and for $D_1^*$, while in the case of ${\tilde D}_2^{*0}$ it could be smaller than $1$ only for a large  mass,  $M({\tilde D}_2^{*0})>3100$ MeV. Instead, a quark model  prediction for the  mass for this meson is $M({\tilde D}_2^{*0})=3035$ MeV  \cite{Di Pierro:2001uu}. Therefore, no quantum number assignment to a meson with a mass of about $2600$ MeV  is able to reproduce the measurement (\ref{new4ratios}).
A possible conclusion is  that there is a  violation of the HQ symmetry in the decays of  $D^{*0}(2600)$.

Let us proceed with the other resonances. Identifying $D^*(2760)$ with $D_3$ and $D(2750)$ with its spin partner $D_2^{\prime *}$, we obtain for the  ratio  (\ref{new4ratios2})
\be
{ {BR}(D^{*0}(2760) \to D^+ \pi^-) \over {BR}(D^{*0}(2750) \to D^{*+} \pi^-)}\Big|_{X^\prime \,{\rm doublet}}=0.660 \pm 0.001 \label{R32A}
\ee
which is  close to the experimental result.
In the hypothesis that $(D(2750),\,D^*(2760))$ fill the $({\tilde D}^\prime_1,\,{\tilde D}_2^*)$ doublet, we obtain
$
\displaystyle{ {BR}(D^{*0}(2760) \to D^+ \pi^-) \over {BR}(D^{*0}(2750) \to D^{*+} \pi^-)}\Big|_{{\tilde T}\, {\rm doublet}}=0.563 \pm 0.001$. Considering the uncertainty in
 (\ref{new4ratios}), both these results agree with the measurement
  \footnote{The ratios  (\ref{R2600},\ref{R32A}) have also been computed in Ref.\cite{Wang:2010ydc}.   The result for the ratio in Eq.(\ref{R2600}) agrees with ours, while
for Eq.(\ref{R32A}) the value $0.80$ has been obtained. The  case of $D^{*0}(2760)$ belonging to the doublet $\tilde T$   has  not been considered in that study.}.

The last  resonance to discuss  in Table \ref{ratios-charm} is $D_{sJ}(2860)$,  for which  we support  the identification  with $D_{s3}$.  However, the ratios in Table \ref{ratios-charm} do not compare favorably with the measurement in Eq.(\ref{exp1})  \cite{Colangelo:2006rq},   and this requires an explanation.
In Section \ref{section:masses} we have predicted that the spin partner of $D_{sJ}(2860)$ has  mass  $M(D_{s2}^{* \prime})=2851 \pm 7$ MeV. Hence, these two states are very close to each other, and the experimental resolution in the common  $D^*K$  decay channel could be difficult. A possible consequence is that the measurement   in  Eq.(\ref{exp1}), which  at first sight is attributed only to $D_{sJ}(2860)$, might be contaminated by the decay $D_{s2}^{* \prime} \to D^*K$, and  what is actually measured is the number of  final $D^{(*)}K$ pairs produced  from both the states:
\be
{\bar R}(2860)= {\Gamma(D_{sJ}(2860) \to D^*K) +\Gamma(D_{s2}^{* \prime}(2851) \to D^*K)
\over \Gamma(D_{sJ}(2860) \to DK)} \,\,\,\, . \label{Rbar2860}  \ee
The prediction for this ratio, considering the two contributions, is
\be
{\bar R}(2860)=0.99 \pm 0.05 \label{Rbar-th}
\ee
 which  agrees with the datum  Eq.(\ref{exp1})\footnote{In \cite{vanBeveren:2009jq} it was proposed that two  overlapping structures with $J^P=0^+$ and $J^P=2^+$  exist in the mass range
 about $2860$ MeV, identified with the n=2 scalar  and tensor $c{\bar s}$ states, respectively.}.

The various ratios of decay rates    for the beauty mesons can be predicted, using
 the observed masses of  the two $2^+$ states in the $s_\ell^P=3/2^+$ doublet and the predicted masses  in Table \ref{bmasses} for the other states.
 The results are  collected in Table \ref{ratios-charm}.
 The last line in the Table corresponds to the assignment of  $D_{sJ}(2860)$ and $D^*(2760)$   to the $X^\prime$, $s_\ell^P=5/2^-$ doublet;  if these two states belong to the $\tilde T$ doublet, for  the corresponding beauty mesons we  predict  $R_\pi=1.15 \pm 0.03$ for ${\tilde B}_2^*$, and $R_K=1.06 \pm  0.01 $, $R_\eta=0.160 \pm 0.003$, $R_\eta^*=0.135 \pm 0.004 $ for ${\tilde B}_{s2}^*$.  Therefore,
 the ratios $R_\pi$ and $R_K$ can be used to distinguish between the two assignments.

\section{Strong coupling constants and  decay widths}
The measurements of the meson  widths allow us to determine the effective coupling constants. Since for mesons in the same doublet one should obtain  the same result in the HQ limit,  a  comparison among the results  tests the  heavy quark symmetry and  the  quantum number assignment  to the decaying state.
From the obtained  coupling constants,   predictions for  not yet observed states follow.

\begin{itemize}

\item{$g$}

The strong transition among states in the $H$  doublet is governed by the coupling $g$. Actually, only  charmed mesons undergo  real   $D^*_{(s)} \to D \pi(K)$  transitions, the corresponding  beauty modes being kinematically forbidden. The coupling constant plays an important role in  processes in which the  $B^*B\pi$ vertex is involved; at present, there is a single experimental determination of the $D^{*\pm}$ width \cite{Anastassov:2001cw}:  $\Gamma(D^{*\pm})=96 \pm 4 \pm 22$ KeV, corresponding to $g=0.64 \pm 0.075 $,  a value which is  larger than the theoretical results obtained in the HQ limit \cite{g,Colangelo:1995ph,Becirevic:2012zz}.

\item $h$

The coupling constant  $h$  controls the decays $S \to H M$.  We can use data on the members of the $c{\bar q}$  doublet $S$,  with $q=u,d$.  However, the case $q=s$ corresponds to  $(D_{s0}^*(2317),\,D_{s1}^\prime(2460))$ and,  as we have already discussed,  the strong decays of these particles cannot be accounted by the effective Lagrangian (\ref{lag-hprimo}) which describes  isospin conserving modes.
Therefore, we only consider the doublet $(D_0^*(2400),\,D_1^\prime(2430))$.
The values of $h$ obtained from their  widths  in Table \ref{charm-pdg}, and using  the rates in Eqs.(\ref{0+Sto0-0-}) and  (\ref{1+Sto1-0-}), are:
$h = 0.61 \pm 0.07$ from $D_{0}^{*0}(2400)$, $h = 0.50 \pm 0.06$ from $D_{0}^{*\pm}(2400)$ and $h = 0.8 \pm 0.2$ from $D_{1}^{\prime 0}(2430)$. The  weighted average is
\be
 h = 0.56 \pm 0.04 \,\,. \label{h-trovato}
 \ee
This result  nicely agrees with the QCD sum rule outcome Ref.\cite{Colangelo:1995ph} and  with the lattice QCD determination in Ref.\cite{Becirevic:2012zz}.
For beauty mesons, using the  predicted masses together with (\ref{h-trovato}),  we obtain the widths  quoted in  Table \ref{bmasses}.

\item{$h^\prime$}

The  $T \to H M$  decays (with  mesons in the doublet $T$ having $n=1$) are  described by the coupling constant $h^\prime$.
Using the  widths  in Table \ref{charm-pdg} together with   Eqs.(\ref{1+Tto1-0-}), (\ref{2+Tto0-0-}) and (\ref{2+Tto1-0-}), we obtain
$h^\prime=0.56 \pm 0.03$ (from $D_1^0(2420)$), $h^\prime=0.54 \pm 0.065$ (from $D_1^\pm(2420)$), $h^\prime=0.43 \pm 0.01$ (from $D_2^{*0}(2460)$),  $h^\prime=0.37 \pm 0.03$ (from $D_2^{*\pm}(2460)$) and  $h^\prime=0.48 \pm 0.035$ (from $D_{s2}^{*}(2573)$). The  weighted average is
\be
 h^\prime = 0.43 \pm 0.01 \,\,. \label{hp-trovato}
 \ee
This  translates into a prediction for the full width of $D_{s1}(2536)$: $\Gamma(D_{s1}(2536))=0.305 \pm 0.002$ MeV.  A recent determination provided by BaBar collaboration
\cite{Lees:2011um},  quoted in Table \ref{charm-pdg},
is  larger than our result,  possibly signalling a  mixing with the other axial-vector state $D_{s1}^\prime(2460)$  \cite{:2007dya}.

In the case of beauty, in the $T$ doublet only  the width of the  $B_{2}^{*0}$ meson  has been measured,   with the result $h^\prime=0.36 \pm 0.09$.
Using this value, the predictions in Table  \ref{bmasses} follow  for the other beauty resonances in $T$.
The two beauty-strange states are very narrow, and  they decay to $B^*K$ (the $B_{s1}(5830)$) and to $BK,\,B^*K$ (the $B_{s2}^{*}(5840)$) with a tiny  phase space ($M_{BK}\simeq 5777$ MeV and $M_{B^*K}\simeq 5823$ MeV).

It is interesting to compare the results obtained from charm and beauty sectors for such a coupling constant, which in the  HQ limit  should coincide. Including $\displaystyle{1/m_Q}$ effects, one may write:  $h^\prime(m_Q)=h^\prime_{asymp}\left(1+\displaystyle{a / m_Q} \right)$. Identifying the value  in (\ref{hp-trovato}) with $h^\prime(m_c)$ (and $m_c \simeq 1.35$ GeV) and the above  $h^\prime(m_b)$ (with $m_b \simeq 4.8$ GeV), we find
$h^\prime_{asymp}=0.33$ and $a=0.13$ GeV. The value from the beauty data is  close to the asymptotic one, while  in the case of charm the correction for the coupling is of   ${\CMcal O}(30\%)$.

\item{$\tilde g$}

 The constant $\tilde g$ governs the decays ${\tilde H} \to HM$, with ${\tilde H}$  the doublet comprising  the radial excitations of $H$.
Observed states  that  fit in such a doublet,  with and without  strangeness, are  the two  resonances  $D(2550),\,D^*(2600)$ and the strange one $D_{s1}^{*}(2700)$. From their measured widths we obtain    $\tilde g=0.35 \pm 0.03$ (from $D(2550)$), $\tilde g=0.23 \pm 0.02$ (from $D^*(2600)$), $\tilde g=0.31 \pm 0.04$ (from $D_{s1}^{*}(2700)$), with  weighted average
\be
{\tilde g}=0.28 \pm 0.015 \label{gtilde-trovato}\,\,.
\ee
We can predict  the full width of the spin partner of $D_{s1}^{*}(2700)$ using the mass fixed in Sec. \ref{section:masses}, as reported in Table \ref{cmasses}.
The predictions for the widths of the beauty resonances belonging to $H^\prime$ are collected in Table \ref{bmasses}.

\end{itemize}

According to the interpretation of the remaining states  [of ($D(2750),\,D^*(2760)$), of $D_{sJ}(2860)$ and its spin partner that we denote with $D_{sJ}(2851)$], we could determine one more constant.
However, for such states other strong decay modes besides those considered here are possible.
Kinematically allowed transitions with the emission of a light vector meson are
\bea
D(2750),\,D^*(2760) \to D\rho,\,\, D \omega  \nonumber
\\
D_{sJ}(2851),\,D_{sJ}(2860) \to DK^*  \label{vec2} \,\,.
\eea
These decays are possible both for states filling  the $X^\prime$, $J^P=(2^-,\,3^-)$ doublet, both for mesons  belonging  to  the doublet $\tilde T$ with $J^P=(1^+,\,2^+)$. In the first case these processes occur in $f$- wave, in the second  one in $d$- wave.
The $2^-$ state could decay also in $p$- wave, and  the $1^+$ meson  in $s$ -wave, but at a next-to-leading order in the  HQ expansion.
The strange meson decays in (\ref{vec2}) are severely   phase-space suppressed.

Other kinematically allowed transitions are the decays to a member of one of the excited doublets ${\tilde H},\,S,\,T$ and  a light pseudoscalar meson.
In the case of  the $X^\prime$ doublet, allowed $f$-wave decay modes are
\bea
D(2750) &\to& D_0^*(2400) \pi, \,\, D_2^*(2460) \pi,  \nonumber
\\
D^*(2760) &\to& D_1^\prime(2430) \pi, \,\, D_1(2429) \pi,\,\,D^*_2(2460) \pi \,\,. \label{other2} \eea
Decays to the members of the doublet $\tilde H$, identified with $(D(2550),\,D^*(2600))$ are possible,  however, the available phase space  is almost closed for $D^*(2600)$,  and is  about $60-70$ MeV for  $D(2550)$.

In the case of  the $\tilde T$ doublet, allowed decay modes are
\bea
D(2750) &\to& D_0^*(2400) \pi, \,\,D_1^\prime(2430) \pi, \,\, D_1(2420) \pi,\,\, D_2^*(2460) \pi,  \nonumber
\\
D^*(2760) &\to& D_1^\prime(2430) \pi, \,\, D_1(2420) \pi,\,\,D^*_2(2460) \pi \,\,, \label{other2T} \eea
which proceed in $p$-wave.
It is important to notice that the transitions  $D(2750) \to D_1^\prime(2430) \pi, \,\, D_1(2420) \pi$
are allowed only if $D(2750)$ has  $J^P=1^+$, $n=2$,  the alternative possibility  considered in this paper. Therefore, experimental study of this decay mode is useful to  establish  the correct classification.

All the  listed modes have  small phase space. Therefore,  we can adopt the same strategy used to determine $h$, $h^\prime$, $\tilde g$  to fix the constant that determines the strong decays of $(D(2750),\,D^*(2760))$ and $D_{sJ}(2860)$,  that depends on the doublet in which we place them. The obtained values should be viewed as upper bounds on the couplings, since we   neglect  suppressed decay modes.

In the classification of  $D_{sJ}(2860)$ as the $J^P=3^-$ state  belonging to the $X^\prime$ doublet, the resonances $(D(2750),\,D^*(2760))$  fill the  corresponding non strange doublet. From their mass and width  we obtain the coupling $k=k_1+k_2$: $k=0.58 \pm 0.05$ (from $D(2750)$), $k=0.39 \pm 0.02$ (from $D^*(2760)$), $k=0.41 \pm 0.03$ (from $D_{sJ}(2860)$). The   average is
\be
k=0.42 \pm 0.02 \,\,. \label{k-trovato}
\ee
Using this result we  predict the full widths of the $D_{s2}^{\prime *}$, the spin partner of $D_{sJ}(2860)$, and of the analogous beauty states [keeping in mind that also for beauty other decay modes are  possible,  the analogous of those in Eqs.(\ref{vec2},\ref{other2})].
Using the masses  in Sec. \ref{section:masses} we obtain the results quoted in Tables  \ref{cmasses} and  \ref{bmasses}.

Alternatively, if  $D_{sJ}(2860)$ is the  $n=2$, $J^P=2^+$ state, i.e. the first radial excitation of $D_{s2}^*(2573)$, and  $(D(2750),\,D(2760))$ the  non strange members of the $\tilde T$ doublet of radial excitations of  $T$, we can fix the constant ${\tilde h}^\prime$ governing the decays ${\tilde T} \to H M$. The results are    ${\tilde h}^\prime=0.23 \pm 0.02$ (from $D(2750)$),
 ${\tilde h}^\prime=0.18 \pm 0.01$ (from $D(2760)$),  ${\tilde h}^\prime=0.17 \pm 0.01$ (from $D_{sJ}(2860)$), with   average
 \be
 {\tilde h}^\prime=0.18 \pm 0.01 \,\,. \label{hptilde-trovato}
\ee
In this case, the spin partner of $D_{sJ}(2860)$ is ${\tilde D}_{s1}$  with $J^P=1^+$, and its full  width is $\Gamma({\tilde D}_{s1})=30 \pm 3$ MeV, while for the ${\tilde T}$ beauty states   we  predict:
$
\Gamma({\tilde B}_{1}) = 96 \pm 7 \,\,\, {\rm MeV}$,
$\Gamma({\tilde B}_{2}^*) = 111 \pm 8.5 \,\,\, {\rm MeV}$,
$\Gamma({\tilde B}_{s1}) = 67 \pm 5 \,\,\, {\rm MeV}$ and
$\Gamma({\tilde B}_{s2}^*) = 83 \pm 6.4 \,\,\, {\rm MeV}$
[neglecting  the modes analogous to  (\ref{vec2}) and (\ref{other2T})].

Finally, we quote our results for the widths of the $s_\ell^P=3/2^+$ beauty states:
$\Gamma(B_1)=13.6\pm0.6$ MeV,
$\Gamma(B_{s1})=0.016\pm0.002$ MeV and
$\Gamma(B^*_{s2})=0.9\pm0.1$ MeV.

\section{Conclusions}
The heavy quark symmetry is a powerful tool for the analysis of the properties of hadrons with a single heavy quark. Using  ideas and methods based on this symmetry, we have proposed a classification of all the observed
$c\bar q$ and $b \bar q$ mesons in doublets, as shown in Table \ref{charm}, determining a set of mass parameters from data. Moreover, we have fixed several effective coupling constants governing the strong transitions into the lightest heavy quark doublet.
With these inputs, we have predicted the mass and width of two not yet observed $c \bar s$ resonances,
reported in Table   \ref{cmasses}. The contribution of the new state $D_{sJ}(2851)$ has been advocated  to explain a discrepancy between the observed ratio of $D^* K/D K$ yield  in the invariant mass range around $2850-2870$ MeV and the theoretical result obtained assuming  only the contribution of $D_{sJ}(2860)$  to this observable.  The other main experimental observables in the charm sector are reproduced, with the  exception of the first ratio in (\ref{new4ratios}) for the $D^*(2600)$. Finally, we have predicted the properties of the not yet observed $b \bar q$ mesons:  the
confirmation of such predictions is expected in the very near future from the experiments  at the LHC.

\section*{Acknowledgement}
We thank D. Milanes and A. Palano for discussions.  This work is supported in part by the Italian MIUR Prin 2009.


\begin{thebibliography}{99}

\bibitem{rev0}
For  reviews see: M.~Neubert,
  Phys.\ Rept.\  {\bf 245}, 259 (1994);
  F.~De Fazio,
 in {\it At the
Frontier of Particle Physics/Handbook of QCD}, ed. by M. Shifman
(World Scientific, Singapore, 2001),  page  1671,
arXiv:hep-ph/0010007.

\bibitem{Isgur:1989ed}
  N.~Isgur and M.~B.~Wise,
  Phys.\ Lett.\ B {\bf 237}, 527 (1990).

\bibitem{Isgur:1991wq}
  N.~Isgur and M.~B.~Wise,
  Phys.\ Rev.\ Lett.\  {\bf 66}, 1130 (1991);
  M.~Lu, M.~B.~Wise and N.~Isgur,
  Phys.\ Rev.\  D {\bf 45}, 1553 (1992).

\bibitem{hqet_chir}
  M.~B.~Wise,
  Phys.\ Rev.\ D {\bf 45}, 2188 (1992);
  G.~Burdman and J.~F.~Donoghue,
  Phys.\ Lett.\ B {\bf 280}, 287 (1992);
  P.~L.~Cho,
  Phys.\ Lett.\ B {\bf 285}, 145 (1992);
  Phys.\ Rev.\ D {\bf 46}, 1148 (1992)
  [Erratum-ibid.\ D {\bf 55}, 5851 (1997)];
    R.~Casalbuoni, A.~Deandrea, N.~Di Bartolomeo, R.~Gatto, F.~Feruglio and G.~Nardulli,
  Phys.\ Lett.\ B {\bf 299}, 139 (1993).

\bibitem{wise_book}
For a review see:
  A.~V.~Manohar and M.~B.~Wise,
  Camb.\ Monogr.\ Part.\ Phys.\ Nucl.\ Phys.\ Cosmol.\  {\bf 10}, 1 (2000).

\bibitem{Falk:1995th}
  A.~F.~Falk and T.~Mehen,
  Phys.\ Rev.\ D {\bf 53} (1996) 231.

\bibitem{pdg}
J. Beringer et al. (Particle Data Group),
Phys.\ Rev.\ D {\bf 86}, 010001 (2012) and http://pdg.lbl.gov/.

\bibitem{Anderson:1999wn}
S.~Anderson {\it et al.}  [CLEO Collaboration],
Nucl. Phys.  {\bf A663}, 647 (2000).

\bibitem{Abe:2003zm}
  K.~Abe {\it et al.}  [Belle Collaboration],
  Phys.\ Rev.\ D {\bf 69}, 112002 (2004).

\bibitem{Link:2003bd}
J.~M.~Link {\it et al.}  [FOCUS Collaboration],
Phys. Lett. {\bf B586}, 11  (2004).

\bibitem{Aubert:2003fg}
B.~Aubert {\it et al.}  [BABAR Collaboration],
Phys. Rev. Lett.  {\bf 90}, 242001 (2003);
%
Y.~Mikami {\it et al.} [Belle Collaboration], Phys. Rev. Lett.
{\bf 92}, 012002 (2004);
%
K.~Abe {\it et al.},
Phys.\ Rev.\ Lett.\  {\bf 92}, 012002 (2004);
%
D.~Besson {\it et al.}  [CLEO Collaboration],
Phys.\ Rev.\ D {\bf 68}, 032002 (2003);
%
L.~Benussi  [FOCUS Collaboration],
  Int.\ J.\ Mod.\ Phys.\  A {\bf 20}, 549 (2005);
  %
B.~Aubert {\it et al.}  [BABAR Collaboration],
Phys. Rev.  {\bf D69}, 031101 (2004);
%
P.~Krokovny {\it et al.} [Belle  Collaboration], Phys. Rev. Lett.
{\bf 91}, 262002 (2003); P.~Krokovny,
  AIP Conf.\ Proc.\  {\bf 717}, 475 (2004).

\bibitem{Colangelo:2005hv}
  P.~Colangelo, F.~De Fazio and A.~Ozpineci,
  Phys.\ Rev.\  D {\bf 72}, 074004 (2005).

\bibitem{Browder:2003fk}
  T.~E.~Browder, S.~Pakvasa and A.~A.~Petrov,
  Phys.\ Lett.\  B {\bf 578}, 365 (2004).

\bibitem{Aubert:2006mh}
B.~Aubert {\it et al.}  [BABAR Collaboration],
  Phys. Rev. Lett.  { \bf 97}, 222001 (2006).

\bibitem{Brodzicka:2007aa}
  J.~Brodzicka {\it et al.}  [Belle Collaboration],
  Phys.\ Rev.\ Lett.\  {\bf 100}, 092001 (2008).

\bibitem{LHCDsJ}
  R.~Aaij {\it et al.}  [ LHCb Collaboration],
  arXiv:1207.6016 [hep-ex].

\bibitem{:2009di}
  B.~Aubert {\it et al.}  [BABAR Collaboration],
  Phys.\ Rev.\  D {\bf 80}, 092003 (2009).

\bibitem{Colangelo:2007ds}
  P.~Colangelo, F.~De Fazio, S.~Nicotri and M.~Rizzi,
  Phys.\ Rev.\  D {\bf 77}, 014012 (2008).

\bibitem{Colangelo:2006rq}
  P.~Colangelo, F.~De Fazio and S.~Nicotri,
  Phys.\ Lett.\  B {\bf 642}, 48 (2006).

\bibitem{Di Pierro:2001uu}
  M.~Di Pierro and E.~Eichten,
  Phys.\ Rev.\  D {\bf 64}, 114004 (2001).

\bibitem{Colangelo:2010te}
  P.~Colangelo and F.~De Fazio,
  Phys.\ Rev.\  D {\bf 81}, 094001 (2010).

\bibitem{delAmoSanchez:2010vq}
  P.~del Amo Sanchez {\it et al.}  [The BABAR Collaboration],
  Phys.\ Rev.\  D {\bf 82}, 111101 (2010).

\bibitem{Godfrey:1985xj}
  S.~Godfrey and N.~Isgur,
  Phys.\ Rev.\  D {\bf 32}, 189 (1985).

\bibitem{Akers:1994fz}
  R.~Akers {\it et al.}  [OPAL Collaboration],
  Z.\ Phys.\  C {\bf 66}, 19 (1995).

\bibitem{lep}
  P.~Abreu {\it et al.}  [DELPHI Collaboration],
  Phys.\ Lett.\  B {\bf 345}, 598 (1995);
  D.~Buskulic {\it et al.}  [ALEPH Collaboration],
  Z.\ Phys.\  C {\bf 69}, 393 (1996);
  R.~Barate {\it et al.}  [ALEPH Collaboration],
  Phys.\ Lett.\  B {\bf 425}, 215 (1998);
  M.~Acciarri {\it et al.}  [L3 Collaboration],
  Phys.\ Lett.\  B {\bf 465}, 323 (1999).

\bibitem{Abazov:2007vq}
  V.~M.~Abazov {\it et al.}  [D0 Collaboration],
  Phys.\ Rev.\ Lett.\  {\bf 99}, 172001 (2007).

\bibitem{:2008jn}
  T.~Aaltonen {\it et al.}  [CDF Collaboration],
  Phys.\ Rev.\ Lett.\  {\bf 102}, 102003 (2009).

\bibitem{:2007tr}
  T.~Aaltonen {\it et al.}  [CDF Collaboration],
  Phys.\ Rev.\ Lett.\  {\bf 100}, 082001 (2008).

\bibitem{:2007sna}
  V.~M.~Abazov {\it et al.}  [D0 Collaboration],
  Phys.\ Rev.\ Lett.\  {\bf 100}, 082002 (2008).

\bibitem{Pappagallo-LHCb}
LHCB Collaboration,  LHCb-CONF-2011-053 (2011).

\bibitem{Colangelo:2003vg}
  P.~Colangelo and F.~De Fazio,
  Phys.\ Lett.\  B {\bf 570}, 180 (2003).

\bibitem{Colangelo:2004vu}
  P.~Colangelo, F.~De Fazio and R.~Ferrandes,
  Mod.\ Phys.\ Lett.\  A {\bf 19}, 2083 (2004).

\bibitem{Colangelo:2005gb}
  P.~Colangelo, F.~De Fazio and R.~Ferrandes,
  Phys.\ Lett.\  B {\bf 634}, 235 (2006).

\bibitem{Wang:2010ydc}
  Z.~G.~Wang,
  Phys.\ Rev.\  D {\bf 83}, 014009 (2011).

\bibitem{vanBeveren:2009jq}
  E.~van Beveren and G.~Rupp,
  Phys.\ Rev.\ D {\bf 81}, 118101 (2010).


\bibitem{Anastassov:2001cw}
  A.~Anastassov {\it et al.}  [CLEO Collaboration],
  Phys.\ Rev.\  D {\bf 65}, 032003 (2002).

\bibitem{g}
P.~Colangelo, G.~Nardulli, A.~Deandrea, N.~Di Bartolomeo, R.~Gatto
and F.~Feruglio,
  Phys.\ Lett.\  B {\bf 339}, 151 (1994);
  P. Colangelo, F. De Fazio and G. Nardulli, Phys. Lett. B{\bf 334}, 175 (1994);
V.~M.~Belyaev, V.~M.~Braun, A.~Khodjamirian and R.~Ruckl,
  Phys.\ Rev.\  D {\bf 51}, 6177 (1995);
 D.~Becirevic, B.~Blossier, E.~Chang and B.~Haas,
  Phys.\ Lett.\  B {\bf 679}, 231 (2009).

\bibitem{Colangelo:1995ph}
  P.~Colangelo, F.~De Fazio, G.~Nardulli, N.~Di Bartolomeo and R.~Gatto,
  Phys.\ Rev.\ D {\bf 52} (1995) 6422;
P.~Colangelo and F.~De Fazio,
  Eur.\ Phys.\ J.\ C {\bf 4} (1998) 503.

\bibitem{Becirevic:2012zz}
  D.~Becirevic, E.~Chang and A.~L.~Yaouanc,
  arXiv:1203.0167 [hep-lat].

\bibitem{Lees:2011um}
  J.~P.~Lees {\it et al.}  [The BABAR Collaboration],
  Phys.\ Rev.\  D {\bf 83}, 072003 (2011).

\bibitem{:2007dya}
  V.~Balagura {\it et al.}  [Belle Collaboration],
  Phys.\ Rev.\  D {\bf 77}, 032001 (2008).

\end{thebibliography}
\end{document}